\documentclass[fleqn,usenatbib]{mnras}

\usepackage{newtxtext,newtxmath}
\usepackage[T1]{fontenc}

\DeclareRobustCommand{\VAN}[3]{#2}
\let\VANthebibliography\thebibliography
\def\thebibliography{\DeclareRobustCommand{\VAN}[3]{##3}\VANthebibliography}

\usepackage{subcaption}
\usepackage{hyperref}
\hypersetup{
  colorlinks=true,
  linkcolor=blue,
  citecolor=blue,
  urlcolor=blue,
}

\usepackage{graphicx}	% Including figure files
\usepackage{amsmath}	% Advanced maths commands
\usepackage{caption}
\usepackage{capt-of}

\usepackage{booktabs}
\usepackage[table,xcdraw]{xcolor}
\usepackage{longtable}

\title[Abundances in the Pleiades Open Cluster]{Chemical Abundances for a Sample of FGK dwarfs in the Pleiades Open Cluster from APOGEE}

\author[V. Grilo et al.]{
Vinicius Grilo,$^{1}$\thanks{E-mail: vinicius.grilo@academico.ufs.br}
Diogo Souto,$^{1}$
Katia Cunha,$^{3,2}$
Rafael Guerço,$^{4,2}$
Rodrigo Vieira,$^{1}$
Verne Smith,$^{5}$
\newauthor
Deusalete Vilar,$^{1}$
Anderson Andrade,$^{1}$
Fabio Wanderley,$^{2}$
Simone Daflon,$^{2}$
João Victor Sales Silva$^{2}$
\\
$^{1}$Departamento de F\'isica, Universidade Federal de Sergipe, Av. Marcelo Deda Chagas, S/N, 49107-230 S\~ao Crist\'ov\~ao, SE, Brazil\\
$^{2}$Observatório Nacional/MCTIC, R. Gen. José Cristino, 77,  20921-400, Rio de Janeiro, Brazil\\
$^{3}$Steward Observatory, University of Arizona, 933 North Cherry Avenue, Tucson, AZ 85721-0065, USA\\
$^{4}$Instituto de Astronom\'ia, Universidad Cat\'olica del Norte, Av. Angamos 0610, Antofagasta, Chile \\
$^{5}$NSF’s NOIRLab, 950 N. Cherry Ave. Tucson, AZ 85719 USA\\
}

\date{Accepted XXX. Received YYY; in original form ZZZ}

\pubyear{2015}

\begin{document}
\label{firstpage}
\pagerange{\pageref{firstpage}--\pageref{lastpage}}
\maketitle

\begin{abstract}
This paper presents chemical abundances of twelve elements (C, Na, Mg, Al, Si, K, Ca, Ti, V, Cr, Mn, and Fe) for 80 FGK dwarfs in the Pleiades open cluster, which span a temperature range of $\sim$2000 K in T$_{\rm eff}$, using the high-resolution (R$\sim$22,500) near-infrared SDSS-IV/APOGEE spectra ($\lambda$1.51–1.69 \micron).
Using a 1D LTE abundance analysis, we determine an overall metallicity of [Fe/H]=+0.03$\pm$0.04 dex, with the elemental ratios [$\alpha$/Fe]=+0.01$\pm$0.05, [odd-z/Fe]=-0.04$\pm$0.08, and [iron peak/Fe]=-0.02$\pm$0.08.
These abundances for the Pleiades are in line with the abundances of other open clusters at similar galactocentric distances as presented in the literature.
Examination of the abundances derived from each individual spectral line revealed that several of the stronger lines displayed trends of decreasing abundance with decreasing $T_{\rm eff}$. The list of spectral lines that yield abundances that are independent of $T_{\rm eff}$ are presented and used for deriving the final abundances.  An investigation into possible causes of the temperature-dependent abundances derived from the stronger lines suggests that the radiative codes and the APOGEE line list we employ may inadequately model van der Waals broadening, in particular in the cooler K dwarfs.
\end{abstract}

\begin{keywords}
infrared: stars --- open clusters and associations: individual --- stars: abundances --- stars: fundamental parameters
\end{keywords}

%%%%%%%%%%%%%%%%%%%%%%%%%%%%%%%%%%%%%%%%%%%%%%%%%%

%%%%%%%%%%%%%%%%% BODY OF PAPER %%%%%%%%%%%%%%%%%%

\section{Introduction} \label{sec:intro}

Open clusters are groups of relatively young, gravitationally bound stars that reside across the Milky Way disk. 
Under the assumption that these open clusters originate from a common progenitor giant molecular cloud (GMC) (\citealt{Krumholz2019}), a uniform and well-mixed chemical distribution composition is expected to result from a single, simultaneous event of star formation.
Consequently, members within these clusters should share not only the same age, galactocentric distance, and initial chemical composition but also differ solely in their initial stellar masses, as indicated by studies such as \cite{DeSilva2006,DeSilva2007}, \cite{Feng2014} and \cite{Bovy2016}. 
This unique scenario makes star clusters one of the most important astrophysical laboratories. They are ideal environments for investigating interactions between stellar chemistry and internal physical processes throughout the stars' evolution, which may contribute to deviations from chemical homogeneity.

The Pleiades open cluster is one of the most extensively studied clusters in the literature, with numerous surveys conducted at various wavelengths to investigate its chemical and dynamic nature (e.g. Gaia (\citealt{Gaia2018}), APOGEE (Apache Point Observatory Galactic Evolution Experiment; \citealt{Majewski2017}) and GALAH (Galactic Archaeology with HERMES; \citealt{DeSilva2015}). 
The availability of Gaia DR2 and DR3 data have elucidated the debate regarding the distance to the Pleiades (\citealt{vanLeeuwen2009,Abramson2018}), inferring that the cluster is situated at a distance of 135.15 $\pm$ 0.43 pc (\citealt{Lodieu2019}), and thus refuting the previous estimate of 120.2 $\pm$ 1.9 pc (\citealt{vanLeeuwen2009}) derived from Hipparcos data.
Although the Pleiades cluster is very close to us, it lies at a position with measurable reddening along the line of sight ($E(B-V) \sim 0.04$ $mag$; \citealt{Odell1994} and \citealt{Gaia2018}). 

This cluster consists of young stars with approximately solar metallicity. Some determinations of metallicity include [Fe/H] values of: [Fe/H] = -0.034 $\pm$ 0.024 (\citealt{Boesgaard1990}); +0.06 $\pm$ 0.05 (\citealt{King2000}); +0.06 $\pm$ 0.02 (\citealt{Gebran2008}); +0.03 $\pm$ 0.05 (\citealt{Funayama2009}); and +0.03 $\pm$ 0.02 (\citealt{Soderblom2009}). 
However, most studies on metallicities and abundances of Pleiades open cluster members focus on solar-like or warmer F-A stars, with a lack of precise abundance analyses in stars covering a wide range of effective temperatures.

The age of the Pleiades open cluster has been debated in the literature often, ranging from 70-130 Myr from determinations using isochrone fitting (\citealt{Mermilliod1981, Vandenberg1984, Mazzei1989, Gossage2018}), while the lithium depletion boundary method (\citealt{Basri1996, Stauffer1998, Barrado2004}) suggests an age of about 112 $\pm$ 5 Myr (\citealt{Dahm2015}).

One of the processes possible to study in stellar clusters is atomic diffusion, a physical process that operates more efficiently in turnoff stars (\citealt{Vauclair1978,Choi2016,Dotter2016,Dotter2017}). Diffusion signatures have been found in stars belonging to globular clusters (NGC 6397 by \citealt{Korn2007}, \citealt{Lind2008}, and \citealt{Nordlander2012}; NGC 6752 by \citealt{Gruyters2013} and \citealt{Gruyters2014}; and M30 by \citealt{Gruyters2016}, \citealt{Gavel2021}, and \citealt{Nordlander2024}) and open clusters (M67 by \citealt{Souto2018,Gao2018,Bertelli2018,Souto2019}, NGC 2420 by \citealt{Semenova2020}, and Coma Berenices by \citealt{Souto2021}). Atomic diffusion is a process dependent on age and metallicity \citep{Michaud2015}. The study of the chemical abundances of Pleiades stellar members offers the possibility to put observational limits to the amount of diffusion taking place at those stars with effective temperatures around 6500 K.

The purpose of this work is to determine the stellar chemical abundances of twelve elements (C, Na, Mg, Al, Si, K, Ca, Ti, V, Cr, Mn, and Fe) using high-resolution APOGEE spectra for F, G, and K main-sequence stars from the Pleiades open cluster, spanning a large range in effective temperatures (from $\sim$ 4500 -- 6800K).
% and to investigate signatures of atomic diffusion in our sample stars. 
This paper is organized as follows: Section \ref{sec:observations} describes the observational data and sample selection; Section \ref{sec:abundance-analysis} details the atmospheric parameters and the methodology used to derive individual abundances; Sections \ref{sec:results} and \ref{sec:discussion}, respectively, present and discuss the main findings; Section \ref{sec:conclusion} offers the concluding remarks of this study.

\section{Observations and sample} \label{sec:observations}

%--------------------------------------------
\begin{figure}
  \includegraphics[width=0.49\textwidth]{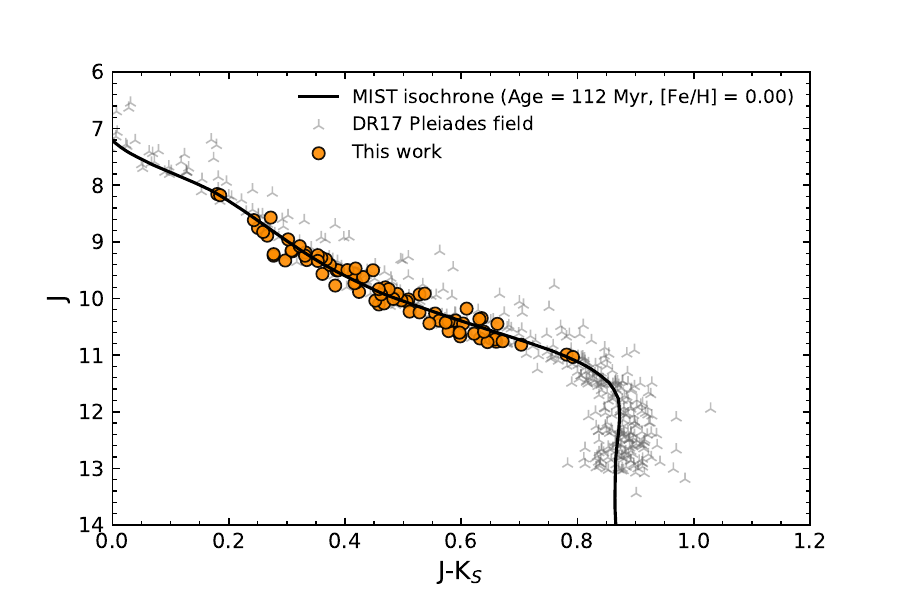}
  \includegraphics[width=0.49\textwidth]{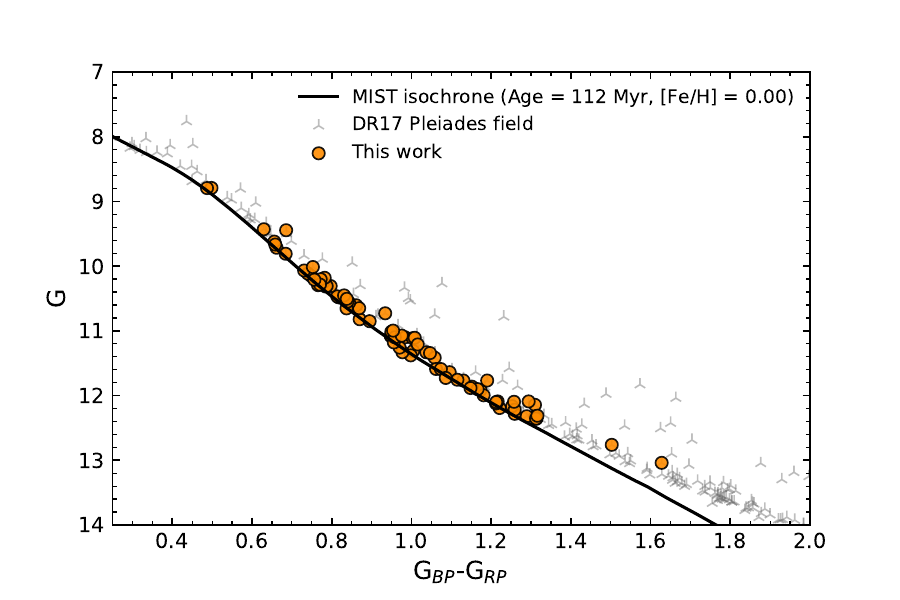}
  \caption{The top and bottom panels show the diagrams (J-K$_{\rm S}$) vs. J and (G$_{\rm BP}$-G$_{\rm RP}$) vs. G for the Pleiades field stars observed by APOGEE (gray inverted triangles), using the reddening corrected photometry from 2MASS and Gaia DR3, respectively. The orange circles represent the stars studied in this work. MIST isochrones are also displayed as solid black lines.}
  \label{fig:CMD} 
\end{figure}
%--------------------------------------------

\subsection{The Apogee Spectra} \label{sec:apogee}

This study is based on data from the SDSS-IV/APOGEE survey. The APOGEE spectrograph was originally commissioned for the Sloan Digital Sky Survey (SDSS) in its third phase (\citealt{Abazajian2005}), continued to be utilized in the fourth phase (\citealt{Adelman-McCarthy2006}) and is currently operational in the fifth phase of SDSS (\citealt{Adelman-McCarthy2007}). The SDSS achieved comprehensive coverage of the sky across both hemispheres, conducting observations at the Las Campanas Observatory (LCO) with the Irénée Du Pont telescope (\citealt{Bowen1973}) and at the Apache Point Observatory (APO) using the Sloan Foundation Telescope (\citealt{Gunn2006}). Both facilities employ telescopes with a 2.5-meter diameter, outfitted with the APOGEE-S and APOGEE-N spectrographs, respectively (\citealt{Wilson2019}).

The APOGEE spectrographs (\citealt{Wilson2010}) are cryogenic, high-resolution (with a spectral resolution R = $\lambda/\delta\lambda$ $\sim 22,500$) spectrographs operating in the H-band, specifically in the near-infrared spectral region between 1.51 – 1.69 $\mu$m. These instruments are multi-object, comprised of a set of fibers, enabling the simultaneous acquisition of stellar spectra from 300 targets.
In the SDSS-V phase, under the Milky Way Mapper survey, the APOGEE spectrographs continue to operate with minor modifications from their SDSS-IV configuration (\citealt{Kollmeier2019}), although the main goal of the survey is to observe main-sequence stars instead of red-giants.

In this work, we utilize spectra from the final public data release of the SDSS-IV/APOGEE, specifically DR17 (\citealt{Abdurrouf2022}). DR17 offers comprehensive coverage of various star clusters (\citealt{Donor2020}), including the Pleiades open cluster, which is the focus of this research. The spectra were processed using the APOGEE pipeline, as detailed in \cite{Nidever2015} with further refinements described in \cite{Holtzman2018} and \cite{Jonsson2020}.

\subsection{The Sample}  \label{sec:sample}

We use the work of \cite{Lodieu2019} as a reference for our membership analysis in the Pleiades. 
\cite{Lodieu2019} provide a revised membership list for three clusters, the Pleiades included, from a five-dimensional analysis ($\alpha, \delta, \pi,\mu_{\alpha}cos\delta,\mu_{\delta}$) using data from the Gaia DR2 catalog (\citealt{Gaia2018}), complemented by comparisons with well known large-scale surveys, such as, the Two Micron All-Sky Survey (2MASS; \citealt{Cutri2003,Skrutskie2006}), the Sloan Digital Sky Survey Data Release 12 (SDSS; \citealt{Abolfathi2018}), the UKIRT Infrared Deep Sky Survey Galactic Clusters Survey (UKIDSS GCS; \citealt{Lawrence2007}), the Wide-field Infrared Survey Explorer (AllWISE; \citealt{Wright2010,Cutri2014}), and the first data release of the Panoramic Survey Telescope and Rapid Response System (PS1; \citealt{Kaiser2002,Chambers2016}). 

The authors obtained a total of 1248 bona-fide members of the Pleiades cluster with a tidal radius of 11.6 pc.
Based on this list, we performed a cross-match with the DR17 APOGEE sample, resulting in 425 stars in common.
We then removed 232 M dwarfs with effective temperatures ($T_{\rm eff}$) $<$ 4100~K from this list because their spectra present significant molecular absorption (particularly from H$_{2}$O), and their study would require an analysis beyond the scope of this work (see discussions in \citealt{Souto2020,Souto2021,Souto2022,Wanderley2023}). 
Also, we removed from the sample 33 stars with $T_{\rm eff}$ $>$ 7000~K because most of their atomic lines were too weak to be precisely studied, resulting in a total of 149 stars.
We note that we used the APOGEE uncalibrated values for the effective temperature in this selection criteria. 
A few stars in this sample were flagged in DR17 with warnings such as SUSPECT\_BROAD\_LINES, SUSPECT\_ROTATION, and PERSIST\_HIGH. To ensure the quality and reliability of the spectra analyzed, we conducted a detailed visual analysis of each spectrum before analysis.

We then removed stars displaying double lines in their spectra and likely to be spectroscopic binaries, stars that were likely binaries based on their positions on the color-magnitude diagram, and those with radial velocity scatter higher than 1.0 km.s$^{-1}$ \citep{ElBadry2018}. We also removed stars having spectra with signal-to-noise (SNR) $<$ 100, and those with vsin($i$) $>$ 30 km.s$^{-1}$, as rotation velocity contributes to the broadening of absorption lines, making abundance determinations more challenging. 
Finally, we note that the stars 2M03491172+2438117 and 2M03440424+2459233 have not been removed from the sample, although they exhibit a radial velocity scatter of 17.7 km.s$^{-1}$ (from 2 APOGEE visits) and 2.9 km.s$^{-1}$ (from 3 APOGEE visits), respectively, suggesting that they are likely members of binary systems. However, we chose to retain them in our sample in order to compare abundance results with \cite{Spina2018} (Section \ref{sec:comp_Optical}).
Our final sample contains 86 stars, covering the effective temperature range roughly between 4500 – 6850~K.

A verification of the radial velocities of the stars in this sample indicated that some of them exhibited radial velocities (RV) that significantly deviated from the expected cluster value of 5.54 $\pm$ 0.10 km.s$^{-1}$ (\citealt{Gaia2018}). These deviations exceeded 3-$\sigma$ of the typical RV uncertainty from APOGEE spectra (1 km.s$^{-1}$).
To refine our membership criteria, we applied the Hierarchical Density-Based Spatial Clustering of Applications with Noise (HDBSCAN; \citealt{Campello2013}) algorithm, using radial velocity as a key indicator. This approach was chosen because \cite{Lodieu2019} did not use radial velocity, and it leverages spectroscopic data from the APOGEE spectra.
The HDBSCAN is a density-based clustering algorithm commonly used in machine learning and data analysis. It builds upon the Density-Based Spatial Clustering of Applications with Noise (DBSCAN, \citealt{Ester1996}) algorithm but introduces a hierarchical approach to cluster extraction, having better performances when compared to DBSCAN and to similar methods like Gaussian Mixture models (\citealt{HuntReffert2021}).

Among the input parameters of the HDBSCAN, the most important are the \textit{minimum number of points} ($m_\mathrm{Pts}$) and the \textit{minimum cluster size} ($m_\mathrm{clSize}$). The $m_\mathrm{Pts}$ is analogous to the epsilon parameter in DBSCAN and defines core and border: points with fewer than $m_\mathrm{Pts}$ neighbors are considered outliers. The $m_\mathrm{clSize}$ sets the minimum number of points required to form a cluster. We followed the recommendation by \cite{Campello2013} and considered the same value for both $m_\mathrm{Pts}$ and $m_\mathrm{clSize}$ parameters. We assumed $m_\mathrm{Pts}$ = $m_\mathrm{clSize}$ = 2. Keeping the values of $m_\mathrm{Pts}$ and $m_\mathrm{clSize}$ as low as possible in HDBSCAN can have several advantages: i) Identifying smaller clusters in the data; ii) $m_\mathrm{Pts}$ low allows the algorithm to classify more points as outliers, which can be useful for identifying rare or anomalous data points that do not belong to any cluster; iii) More flexibility in capturing diverse structures within the data; and iv) Lowering these parameters can encourage the algorithm to explore different density levels in the data, potentially revealing hierarchical structures or nested clusters that might be missed with higher parameter values.

We ran the HDBSCAN in our sample of 86 stars, assuming the stellar spectroscopic radial velocity from DR17 and proper motions ($\mu_{\alpha}cos\delta$, $\mu_{\delta}$) from Gaia DR3 (\citealt{GaiaDR3}). 
We confirmed 80 stars belonging to the Pleiades, which composes our final sample shown in Table \ref{tab:sample}.

In Figure \ref{fig:CMD}, we present color-magnitude diagrams (CMDs) using dereddened data from 2MASS (\citealt{Skrutskie2006}) (J-K$_{S}$ – J) and Gaia DR3 (G$_{BP}$-G$_{RP}$ – G). The selected targets are displayed as orange circles, while the sample of APOGEE targets observed in the Pleiades field are shown as gray inverted triangles. 
For comparison, we also display a MIST isochrone (\citealt{Choi2016, Dotter2016}) with an age of 112 Myr and solar metallicity as a solid black line, corresponding to the age and metallicity of the Pleiades (\citealt{Dahm2015}). 
All magnitudes underwent correction for extinction, employing the mean Pleiades extinction value of A$_{\rm V}$ = 0.12 \citep{stauffer2007_vmag} and the relationships provided by \citet{Wang2019}. 
The selected stars from the Pleiades (orange circles) are unevolved, i.e., they have not yet reached the turnoff point, where stars begin their rapid hydrogen exhaustion phase. The points representing the sample and the displayed MIST isochrone agree well. 

\section{Chemical Abundance Analysis} \label{sec:abundance-analysis}

\subsection{Atmospheric Parameters}  \label{sec:atm_parameters}

The effective temperature values adopted in this study were obtained from APOGEE DR17. We used the raw values from ASPCAP (APOGEE Stellar Parameters and Chemical Abundances Pipeline; \citealt{GarciaPerez2016}), a pipeline developed by the APOGEE team for determining stellar parameters ($T\rm{_{eff}}$, log $g$, [M/H], [C/M], [N/M], [$\alpha$/H], $\xi$) and chemical abundances for over 24 elements in all stars observed by APOGEE. ASPCAP uses the FERRE code (\citealt{Allende2006, Allende2014}), and it compares the spectra observed by APOGEE to a library of synthetic spectra using chi-square minimization ($\chi^2$) and identifies the set of stellar parameters and abundances that provide the best fit to the synthetic spectrum.

The surface gravities (log $g$) for the stars were calculated using the Stefan-Boltzmann equation, expressed as follows:

\begin{align}
	\log g = \log g_{\odot} + \log\bigg(\dfrac{M_{\star}}{M_{\odot}}\bigg) + 4\log\bigg(\dfrac{T_{\rm eff \star}}{T_{\odot}}\bigg)-\dfrac{L_{\rm bol,\star}}{L_{\rm bol,\odot}} 
	\label{eq-stefan-boltzmann}
\end{align}

%The values for the effective temperature were taken from the raw ASPCAP data. 
Information on stellar mass and luminosity was derived through interpolation of a 125 million-year-old isochrone with a metallicity of [Fe/H] = 0.00 dex, representative of the Pleiades cluster, from the MESA Isochrones and Stellar Tracks (MIST; \citealt{Choi2016,Dotter2016}). The solar reference values used were $T_{\text{eff},\odot}$ = 5772 K, $\log g_\odot$ = 4.438 dex, and bolometric magnitude M$_{\text{bol},\odot}$ = 4.75, as recommended by the IAU (\citealt{Prsa2016}).

\onecolumn
{\scriptsize
\renewcommand{\arraystretch}{0.95}
\begin{longtable}{@{}lccccccccccl@{}}
% \tablenum(1)
\caption{Atmospheric Parameters} 
\label{tab:atm-parameters}\\
\hline
2MASS ID           & J      & H      & K      & pm \textit{ra} (mas yr$^{-1}$)           & pm \textit{dec} (mas yr$^{-1}$)            & RV (km s$^{-1}$)   & SNR & \textit{vsin(i)} (km s$^{-1}$) & T$\rm _{eff}$ (K) & $\log $g (cm s$^{-2}$) & $\xi $ (km s$^{-1}$) \\ 
\hline
2M03450528+2342097 & 8.205  & 8.041  & 8.000  & 19.75 $\pm$ 0.02 & -46.40 $\pm$ 0.02 & 6.2   & 641 & 21.0 $\pm$ 1.2          & 6848          & 4.33     & 1.6               \\
2M03445123+2316082 & 8.184  & 8.006  & 7.984  & 20.55 $\pm$ 0.02 & -44.30 $\pm$ 0.02 & 7.1   & 617 & 26.5  $\pm$ 1.9         & 6678          & 4.35     & 1.6               \\
2M03435880+2352578$^{a}$ & 8.600  & 8.420  & 8.308  & 19.88 $\pm$ 0.06 & -44.69 $\pm$ 0.05 & 6.0   & 527 & 14.6 $\pm$ 0.9         & 6577          & 4.37     & 1.6               \\
2M03501766+2522464 & 8.645  & 8.469  & 8.382  & 19.35 $\pm$ 0.03 & -45.86 $\pm$ 0.02 & 6.5   & 684 & 24.5 $\pm$ 1.0         & 6451          & 4.39     & 1.6               \\
2M03475252+2356286 & 8.853  & 8.639  & 8.574  & 19.68 $\pm$ 0.03 & -46.39 $\pm$ 0.02 & 6.7   & 486 & 20.1 $\pm$ 0.0        & 6251          & 4.43     & 1.6               \\
2M03482616+2402544$^{b}$ & 8.985  & 8.719  & 8.663  & 19.67 $\pm$ 0.02 & -46.26 $\pm$ 0.01 & 6.5   & 306 & 16.2 $\pm$ 1.0         & 6206          & 4.44     & 1.6               \\
2M03444075+2449067 & 8.783  & 8.540  & 8.513  & 20.24 $\pm$ 0.03 & -45.45 $\pm$ 0.02 & 4.7   & 581 & 19.4 $\pm$ 0.8         & 6169          & 4.45     & 1.6               \\
2M03394117+2317271 & 8.920  & 8.686  & 8.634  & 21.29 $\pm$ 0.02 & -43.71 $\pm$ 0.01 & 6.3   & 469 & 28.4 $\pm$ 1.4         & 6159          & 4.45     & 1.6               \\
2M03463878+2457346 & 9.105  & 8.851  & 8.763  & 20.61 $\pm$ 0.02 & -46.95 $\pm$ 0.01 & 6.4   & 574 & 21.0 $\pm$ 0.8         & 6095          & 4.46     & 1.6               \\
2M03491172+2438117$^{a,b}$ & 9.246  & 9.028  & 8.949  & 20.68 $\pm$ 0.03 & -44.74 $\pm$ 0.02 & 3.9   & 273 & 12.5 $\pm$ 0.9         & 6055          & 4.47     & 1.6               \\
2M03504007+2355590 & 9.181  & 8.941  & 8.853  & 18.71 $\pm$ 0.03 & -44.36 $\pm$ 0.02 & 6.7   & 559 & 24.6 $\pm$ 0.9         & 6030          & 4.48     & 1.6               \\
2M03385686+2434112$^{a}$ & 9.358  & 9.053  & 9.041  & 20.16 $\pm$ 0.02 & -43.46 $\pm$ 0.02 & 5.9   & 530 & 11.2 $\pm$ 0.8         & 6029          & 4.48     & 1.6               \\
2M03462267+2434126$^{b}$ & 9.274  & 8.994  & 8.923  & 19.68 $\pm$ 0.02 & -44.99 $\pm$ 0.01 & 6.9   & 583 & 12.7 $\pm$ 0.7         & 6024          & 4.48     & 1.6               \\
2M03445639+2425574$^{b}$ & 9.218  & 8.945  & 8.866  & 19.70 $\pm$ 0.03 & -46.14 $\pm$ 0.02 & 5.3   & 587 & 18.6 $\pm$ 0.7         & 6020          & 4.48     & 1.6               \\
2M03462735+2508080$^{b}$ & 9.366  & 9.063  & 8.993  & 20.87 $\pm$ 0.02 & -45.54 $\pm$ 0.02 & 6.6   & 497 & 12.5 $\pm$ 0.6        & 6006          & 4.48     & 1.6               \\
2M03483451+2326053$^{b}$ & 9.197  & 8.980  & 8.868  & 19.37 $\pm$ 0.02 & -45.08 $\pm$ 0.01 & 6.4   & 478 & 15.4 $\pm$ 0.8         & 5979          & 4.49     & 1.6               \\
2M03481712+2353253$^{b}$ & 9.264  & 8.976  & 8.891  & 17.89 $\pm$ 0.02 & -45.50 $\pm$ 0.01 & 6.2   & 284 & 14.1 $\pm$ 1.0         & 5966          & 4.49     & 1.6               \\
2M03464706+2254525$^{b}$ & 9.307  & 9.032  & 8.928  & 18.76 $\pm$ 0.02 & -45.00 $\pm$ 0.02 & 6.8   & 470 & 17.3 $\pm$ 0.8         & 5897          & 4.50     & 1.2               \\
2M03465491+2447468$^{a,b}$ & 9.274  & 9.005  & 8.977  & 19.30 $\pm$ 0.02 & -46.34 $\pm$ 0.01 & 6.0   & 483 & 8.7 $\pm$ 0.5          & 5869          & 4.51     & 1.2               \\
2M03465373+2335009$^{b}$ & 9.350  & 9.092  & 8.996  & 20.80 $\pm$ 0.02 & -45.15 $\pm$ 0.01 & 6.6   & 389 & 8.7 $\pm$ 0.6          & 5857          & 4.51     & 1.2               \\
2M03441391+2446457 & 9.530  & 9.172  & 9.062  & 20.64 $\pm$ 0.02 & -45.24 $\pm$ 0.01 & 6.5   & 432 & 8.9 $\pm$ 0.7          & 5842          & 4.51     & 1.2               \\
2M03474044+2421525$^{b}$ & 9.340  & 9.071  & 8.953  & 19.32 $\pm$ 0.03 & -45.46 $\pm$ 0.02 & 5.5   & 538 & 15.9 $\pm$ 0.7          & 5800          & 4.52     & 1.2               \\
2M03450400+2515282$^{a,b}$ & 9.435  & 9.146  & 9.041  & 18.65 $\pm$ 0.02 & -45.41 $\pm$ 0.02 & 5.4   & 441 & 12.6 $\pm$ 0.7          & 5796          & 4.52     & 1.2               \\
2M03440424+2459233$^{a}$ & 9.499  & 9.168  & 9.061  & 18.69 $\pm$ 0.23 & -47.79 $\pm$ 0.13 & 3.0   & 536 & 8.9 $\pm$ 0.7           & 5755          & 4.53     & 1.2               \\
2M03440059+2332382 & 9.650  & 9.274  & 9.199  & 19.56 $\pm$ 0.02 & -44.75 $\pm$ 0.01 & 7.4   & 333 & 12.7 $\pm$ 0.6          & 5706          & 4.54     & 1.2               \\
2M03433195+2340266 & 9.527  & 9.192  & 9.103  & 19.92 $\pm$ 0.07 & -45.32 $\pm$ 0.05 & 5.5   & 323 & 12.0 $\pm$ 0.8          & 5688          & 4.54     & 1.2               \\
2M03493312+2347435 & 9.595  & 9.294  & 9.214  & 19.92 $\pm$ 0.02 & -43.75 $\pm$ 0.01 & 7.3   & 396 & 6.4 $\pm$ 0.7           & 5688          & 4.54     & 1.2               \\
2M03405042+2325064 & 9.531  & 9.192  & 9.123  & 21.70 $\pm$ 0.02 & -45.50 $\pm$ 0.01 & 5.8   & 386 & 11.7 $\pm$ 0.7          & 5656          & 4.55     & 1.2               \\
2M03433772+2332096 & 9.529  & 9.191  & 9.124  & 19.53 $\pm$ 0.02 & -45.29 $\pm$ 0.01 & 6.4   & 433 & 12.3 $\pm$ 0.5          & 5641          & 4.55     & 1.2               \\
2M03481769+2502523 & 9.593  & 9.256  & 9.158  & 19.52 $\pm$ 0.02 & -47.06 $\pm$ 0.02 & 6.3   & 479 & 12.2 $\pm$ 0.6          & 5577          & 4.56     & 1.2               \\
2M03403436+2340574 & 9.585  & 9.241  & 9.148  & 20.55 $\pm$ 0.02 & -45.63 $\pm$ 0.01 & 5.7   & 347 & 10.8 $\pm$ 0.7          & 5566          & 4.56     & 1.2               \\
2M03474811+2313053 & 9.760  & 9.436  & 9.324  & 20.05 $\pm$ 0.03 & -45.85 $\pm$ 0.02 & 7.6   & 416 & 4.8 $\pm$ 1.0           & 5506         & 4.57     & 1.2               \\
2M03502089+2428003 & 9.746  & 9.397  & 9.307  & 20.35 $\pm$ 0.03 & -45.48 $\pm$ 0.02 & 5.6   & 455 & 8.7 $\pm$ 0.6           & 5501          & 4.57     & 1.0               \\
2M03454184+2425534 & 9.801  & 9.528  & 9.398  & 20.51 $\pm$ 0.03 & -44.71 $\pm$ 0.02 & 7.3   & 350 & 7.1 $\pm$ 0.8           & 5471          & 4.57     & 1.0               \\
2M03502130+2305470 & 9.863  & 9.512  & 9.385  & 19.61 $\pm$ 0.03 & -46.56 $\pm$ 0.02 & 7.4   & 393 & 6.8 $\pm$ 0.8           & 5406          & 4.58     & 1.0               \\
2M03444398+2413523 & 9.913  & 9.566  & 9.469  & 21.47 $\pm$ 0.03 & -46.45 $\pm$ 0.02 & 5.5   & 412 & 9.1 $\pm$ 0.8           & 5370          & 4.59     & 1.0               \\
2M03392780+2353420 & 9.957  & 9.556  & 9.475  & 21.16 $\pm$ 0.03 & -45.35 $\pm$ 0.02 & 5.0   & 300 & 5.5 $\pm$ 1.1           & 5368          & 4.59     & 1.0               \\
2M03433440+2345429 & 9.865  & 9.476  & 9.370  & 22.97 $\pm$ 0.04 & -47.12 $\pm$ 0.03 & 5.3   & 334 & 6.3 $\pm$ 1.0           & 5360          & 4.59     & 1.0               \\
2M03435070+2414508 & 9.833  & 9.429  & 9.344  & 21.61 $\pm$ 0.03 & -44.80 $\pm$ 0.02 & 6.4   & 413 & 10.7 $\pm$ 0.6          & 5323          & 4.59     & 1.0               \\
2M03470141+2329419 & 10.037 & 9.637  & 9.534  & 19.07 $\pm$ 0.02 & -45.02 $\pm$ 0.02 & 6.3   & 248 & 8.2 $\pm$ 0.9           & 5271          & 4.60     & 1.0               \\
2M03492873+2342440 & 9.914  & 9.543  & 9.428  & 19.94 $\pm$ 0.03 & -47.45 $\pm$ 0.02 & 6.1   & 402 & 8.2 $\pm$ 0.6           & 5270          & 4.60     & 1.0               \\
2M03495035+2342202 & 10.067 & 9.699  & 9.595  & 19.59 $\pm$ 0.03 & -44.02 $\pm$ 0.02 & 7.1   & 346 & 6.0 $\pm$ 0.9           & 5264          & 4.60     & 1.0               \\
2M03505508+2411508 & 10.116 & 9.740  & 9.629  & 18.86 $\pm$ 0.02 & -43.39 $\pm$ 0.02 & 6.2   & 393 & 9.0 $\pm$ 0.8           & 5216          & 4.60     & 1.0               \\
2M03532369+2403542 & 9.949  & 9.564  & 9.439  & 19.43 $\pm$ 0.03 & -45.30 $\pm$ 0.02 & 6.9   & 216 & 13.7 $\pm$ 1.1          & 5183          & 4.61     & 1.0               \\
2M03490232+2315088 & 10.135 & 9.758  & 9.657  & 19.73 $\pm$ 0.02 & -44.43 $\pm$ 0.02 & 6.1   & 304 & 6.8 $\pm$ 1.0           & 5159          & 4.61     & 1.0               \\
2M03461174+2437203 & 10.062 & 9.629  & 9.545  & 21.20 $\pm$ 0.03 & -44.70 $\pm$ 0.02 & 6.9   & 397 & 8.3 $\pm$ 0.8           & 5106          & 4.61     & 1.0               \\
2M03444317+2552319 & 10.031 & 9.669  & 9.520  & 19.77 $\pm$ 0.03 & -45.90 $\pm$ 0.02 & 4.1   & 328 & 6.2 $\pm$ 0.9           & 5103          & 4.61     & 1.0               \\
2M03440484+2416318 & 9.943  & 9.494  & 9.386  & 20.13 $\pm$ 0.02 & -45.36 $\pm$ 0.02 & 5.3   & 205 & 10.8 $\pm$ 0.9          & 5095          & 4.61     & 1.0               \\
2M03573331+2403114 & 10.454 & 9.965  & 9.861  & 18.76 $\pm$ 0.02 & -45.31 $\pm$ 0.01 & 6.3   & 253 & 2.9 $\pm$ 1.4           & 5016          & 4.62     & 1.0               \\
2M03430293+2440110 & 10.275 & 9.808  & 9.727  & 19.45 $\pm$ 0.02 & -46.25 $\pm$ 0.01 & 5.6   & 366 & 8.2 $\pm$ 0.7           & 5015          & 4.62     & 1.0               \\
2M03450326+2350219 & 10.261 & 9.835  & 9.730  & 20.34 $\pm$ 0.02 & -45.77 $\pm$ 0.01 & 4.5   & 158 & 6.5 $\pm$ 1.4           & 5008          & 4.62     & 1.0               \\
2M03513903+2245010 & 10.435 & 10.013 & 9.837  & 19.60 $\pm$ 0.02 & -45.60 $\pm$ 0.01 & 6.8   & 287 & 3.4 $\pm$ 1.3           & 4997          & 4.62     & 1.0               \\
2M03363030+2400440 & 10.468 & 10.026 & 9.903  & 19.76 $\pm$ 0.02 & -43.17 $\pm$ 0.01 & 6.0   & 237 & 4.4 $\pm$ 1.5           & 4936          & 4.63     & 1.0               \\
2M03403072+2429143 & 10.082 & 9.659  & 9.556  & 21.82 $\pm$ 0.04 & -44.17 $\pm$ 0.03 & 4.0   & 332 & 6.6 $\pm$ 1.2           & 4914          & 4.63     & 1.0               \\
2M03444394+2529574 & 10.391 & 9.863  & 9.740  & 20.32 $\pm$ 0.02 & -43.83 $\pm$ 0.01 & 5.9   & 276 & 7.8 $\pm$ 1.0           & 4906          & 4.63     & 1.0               \\ 
2M03470678+2342546 & 10.421 & 9.995  & 9.839  & 19.43 $\pm$ 0.02 & -44.41 $\pm$ 0.01 & 6.6   & 149 & 3.3 $\pm$ 1.5           & 4826          & 4.64     & 1.0               \\
2M03471480+2522186 & 10.046 & 9.515  & 9.517  & 19.56 $\pm$ 0.02 & -45.94 $\pm$ 0.02 & 5.4   & 364 & 10.4 $\pm$ 0.7          & 4813          & 4.64     & 1.0               \\
2M03434901+2543466 & 10.294 & 9.834  & 9.719  & 21.27 $\pm$ 0.02 & -47.77 $\pm$ 0.01 & 5.2   & 309 & 10.0 $\pm$ 0.9          & 4800          & 4.64     & 1.0               \\
2M03511685+2349357 & 10.616 & 10.077 & 9.957  & 19.84 $\pm$ 0.02 & -46.76 $\pm$ 0.01 & 6.5   & 99  & 6.7 $\pm$ 2.2           & 4777          & 4.64     & 1.0               \\
2M03441120+2322455 & 9.956  & 9.512  & 9.408  & 22.04 $\pm$ 0.03 & -46.64 $\pm$ 0.02 & 6.8   & 374 & 18.2 $\pm$ 0.7          & 4774          & 4.64     & 1.0               \\
2M03505143+2319447 & 10.632 & 10.108 & 10.015 & 18.89 $\pm$ 0.02 & -46.33 $\pm$ 0.01 & 7.1   & 129 & 4.7 $\pm$ 1.9           & 4736          & 4.65     & 1.0               \\
2M03452219+2328182 & 10.471 & 9.959  & 9.848  & 20.22 $\pm$ 0.02 & -44.65 $\pm$ 0.01 & 5.9   & 167 & 7.9 $\pm$ 1.2           & 4733          & 4.65     & 1.0               \\
2M03404256+2542197 & 10.799 & 10.240 & 10.134 & 20.79 $\pm$ 0.02 & -45.87 $\pm$ 0.01 & 5.0   & 262 & 8.0 $\pm$ 1.1           & 4667          & 4.65     & 1.0               \\
2M03440509+2529017 & 10.605 & 10.087 & 10.007 & 20.77 $\pm$ 0.02 & -45.66 $\pm$ 0.01 & 5.3   & 301 & 8.0 $\pm$ 1.0           & 4667          & 4.65     & 1.0               \\
2M03471352+2342515 & 10.624 & 10.148 & 10.009 & 19.73 $\pm$ 0.02 & -45.61 $\pm$ 0.01 & 6.6   & 156 & 6.2 $\pm$ 1.4           & 4638          & 4.65     & 1.0               \\
2M03463938+2401468 & 10.454 & 9.986  & 9.859  & 20.27 $\pm$ 0.02 & -46.23 $\pm$ 0.01 & 5.9   & 171 & 8.2 $\pm$ 1.1           & 4634          & 4.65     & 1.0               \\
2M03420470+2553091 & 10.784 & 10.183 & 10.093 & 21.06 $\pm$ 0.02 & -46.90 $\pm$ 0.01 & 4.7   & 256 & 9.6 $\pm$ 0.9           & 4623          & 4.65     & 1.0               \\
2M03555603+2334021 & 10.651 & 10.119 & 10.009 & 19.14 $\pm$ 0.02 & -46.08 $\pm$ 0.01 & 7.2   & 226 & 8.7 $\pm$ 0.9           & 4610          & 4.66     & 1.0               \\
2M03415906+2555153 & 10.752 & 10.167 & 10.074 & 20.58 $\pm$ 0.02 & -45.14 $\pm$ 0.01 & 5.4   & 252 & 7.6 $\pm$ 1.3           & 4602          & 4.66     & 1.0               \\
2M03422759+2502492 & 10.736 & 10.200 & 10.065 & 19.37 $\pm$ 0.02 & -43.27 $\pm$ 0.01 & 5.9   & 308 & 8.6 $\pm$ 0.7           & 4599          & 4.66     & 1.0               \\
2M03460649+2434027 & 10.793 & 10.236 & 10.113 & 19.75 $\pm$ 0.02 & -46.62 $\pm$ 0.01 & 5.5   & 271 & 8.3 $\pm$ 0.8           & 4598          & 4.66     & 1.0               \\
2M03503457+2430281 & 10.698 & 10.197 & 10.080 & 19.84 $\pm$ 0.02 & -45.04 $\pm$ 0.02 & 6.4   & 307 & 10.7 $\pm$ 0.8          & 4571          & 4.66     & 1.0               \\
2M03401202+2538321 & 10.208 & 9.757  & 9.579  & 18.35 $\pm$ 0.04 & -47.55 $\pm$ 0.03 & 3.8   & 363 & 15.0 $\pm$ 0.8          & 4570          & 4.66     & 1.0               \\
2M03513927+2432561 & 10.418 & 9.921  & 9.808  & 21.23 $\pm$ 0.02 & -46.06 $\pm$ 0.01 & 6.4   & 174 & 10.4 $\pm$ 1.0          & 4552          & 4.66     & 1.0               \\
2M03540892+2420011 & 10.376 & 9.845  & 9.721  & 16.63 $\pm$ 0.03 & -46.42 $\pm$ 0.02 & 7.5   & 323 & 6.3 $\pm$ 1.3           & 4536          & 4.66     & 1.0               \\
2M03413245+2309422 & 10.845 & 10.255 & 10.122 & 21.49 $\pm$ 0.03 & -46.00 $\pm$ 0.02 & 5.4   & 93  & 6.6 $\pm$ 2.1           & 4531          & 4.66     & 1.0               \\
2M03452957+2345379 & 11.063 & 10.417 & 10.251 & 19.07 $\pm$ 0.02 & -45.76 $\pm$ 0.01 & 6.3   & 254 & 6.3 $\pm$ 1.3           & 4525         & 4.66     & 1.0               \\
2M03405126+2335543 & 10.735 & 10.209 & 10.083 & 21.13 $\pm$ 0.02 & -45.76 $\pm$ 0.01 & 5.7   & 124 & 8.1 $\pm$ 1.7           & 4521          & 4.66     & 1.0               \\
2M03432662+2459395 & 10.477 & 9.956  & 9.795  & 19.04 $\pm$ 0.05 & -44.59 $\pm$ 0.03 & 5.6   & 294 & 9.9 $\pm$ 0.9           & 4514          & 4.66     & 1.0               \\
2M03455048+2352262 & 11.020 & 10.382 & 10.219 & 19.88 $\pm$ 0.02 & -44.65 $\pm$ 0.01 & 5.0   & 217 & 7.6 $\pm$ 1.2           & 4502          & 4.66     & 1.0               \\ 
\hline
\label{tab:sample}
% \footnotesize
% \textit{Notes:} \\
$^a$ Stars studied by & \cite{Spina2018}.\\
$^b$ Stars studied by & \cite{Soderblom2009}.\\
\end{longtable}
}
\twocolumn

The microturbulence velocity ($\xi$) was derived following the same procedure as described by \cite{Smith2013} (see also \citealt{Souto2016}). It consists of varying the $\xi$ from 0.5 to 3.0 in steps of 0.5 km.s$^{-1}$. We adopted the value that results in the lowest dispersion in the abundances of Fe I lines as the microturbulence velocity for the star. 

\subsection{Metallicities and v sin($i$) determinations using MCMC}  \label{sec:MCMC}

We computed spectral syntheses using the radiative transfer code TurboSpectrum (\citealt{Alvarez1998, Plez2012}) and one-dimensional (1-D) local thermodynamic equilibrium (LTE) and plane-parallel atmospheric models interpolated from the MARCS grid (\citealt{Gustafsson2008}). The synthesis uses the APOGEE spectral line list for the H-band ($\lambda$ = 15,000-17,000 \AA), which was incorporated into SDSS-V DR16 as outlined by \cite{Smith2021}. This line list includes critical transitions for analyzing our sample spectra.

The first step in this analysis was to use the Markov chain Monte Carlo (MCMC) technique to analyze a selection of Fe I lines in the APOGEE spectra, 
employing a two-dimensional grid of synthetic spectra. The key parameters in our analysis were the iron abundance (A(Fe)) and the projected rotational velocity (vsin($i$)). To build this synthetic grid, we maintained fixed values for each target's effective temperature, surface gravity, and microturbulence, as specified in Table \ref{tab:atm-parameters}. The grid varies A(Fe) from 6.90 to 7.70 in 0.01 dex increments and vsin($i$) from 1.0 to 40.0 km.s$^{-1}$ in 1 km.s$^{-1}$ steps.

For each star, we generated a comprehensive grid comprising 1640 spectral syntheses.
To correct the fiber-to-fiber and wavelength-dependent line spread function (LSF) variations exhibited in APOGEE spectra, we convolved the synthetic spectra with the LSF to match the APOGEE resolution (\citealt{Nidever2015,Wilson2019}).

We calculated chi-squared values for each spectral line considering line center adjustments of $\pm$5\AA{} around all well-defined Fe I lines within the APOGEE spectral range. Our line selection is based on the criteria described in \cite{Souto2018}, with exclusions made for weak lines or those impacted by spectral reduction issues.

The posterior probability of the parameters was sampled using \texttt{emcee}\footnote{Available online under the MIT License: \url{https:// github.com/ dfm/ emcee}.} \citep{foreman2013}, a Python implementation of the Affine Invariant MCMC Ensemble sampler \citep{goodman2010}. We adopted a likelihood function proportional to $\exp(-\chi_\mathrm{tot}^2/2)$, where  $\chi_\mathrm{tot}^2$ corresponds the the sum of the individual line $\chi^2$'s. For intermediate grid values tested during the \texttt{emcee} sampling, we interpolated the $\chi_\mathrm{tot}$ values using \texttt{griddata} routine, from Python Scipy\footnote{For more details on the SCIPY package, refer to \citet{virtanen2020}.}. For each simulation, we used $50$~walkers (random-walk samplers) with $50$~steps in the initial phase (burn-in) and $100$ steps in the final sampling phase (starting from the last state of the burn-in chain).
These values were selected because a convergence plateau (seeing a trace plot) for both A(Fe) and vsin($i$) is observed between 50 and 100 steps in the analysis.

\subsection{Abundance determinations using BACCHUS}  \label{sec:abundances}
 
Abundances of C, Na, Mg, Al, Si, K, Ca, Ti, V, Cr, Mn, and Fe were calculated using the BACCHUS wrapper (\citealt{Masseron2016}), in semi-automatic mode, which employs the radiative transfer code TurboSpectrum (\citealt{Alvarez1998, Plez2012}) to generate spectral syntheses that are directly compared to the observed stellar spectra. We convolved the synthetic spectra assuming the APOGEE resolution and the derived vsin($i$) values, and we determined the best-fit syntheses from varying the elemental abundances and using the BACCHUS code to perform chi-squared minimizations between observed and model spectra. The best fits obtained were then confirmed
from a visual inspection to verify the quality of the fits.

%--------------------------------------------
\begin{figure}
  \includegraphics[width=0.48\textwidth]{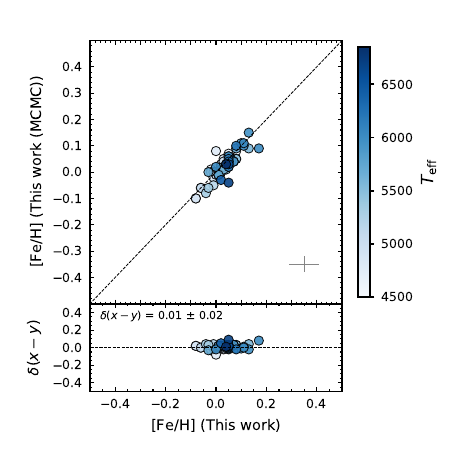}
  \includegraphics[width=0.48\textwidth]{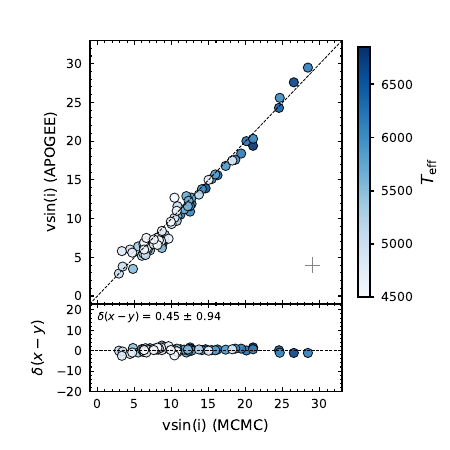}  
  \caption{Upper panel: The [Fe/H] $vs.$ [Fe/H] distribution with this work's results from the semi-automatic BACCHUS abundance analysis on the x-axis and from MCMC on the y-axis. The subpanel shows the residual (x–y) diagram, where a mean value difference with the respective standard deviation of the mean is defined as $\delta$. A color bar is used to represent the effective temperatures of the stars. Lower panel: same format as the upper panel, but comparing vsin($i$) values measured by our MCMC analysis on the x-axis and from APOGEE on the y-axis. A typical error bar is presented at the bottom right of the figure.}
  \label{fig:vsinia}
\end{figure}
%--------------------------------------------

\subsection{Excluding spectral lines with spurious trends}  \label{sec:removing_lines}

To avoid inconsistent abundance results, we excluded those lines whose abundances exhibited strong correlations with $T_{\rm eff}$.
From an initial list of atomic lines presented in \cite{Souto2018}, we removed two C I lines ($\lambda$15784.7\AA, $\lambda$16005\AA), three Mg I lines ($\lambda$15740.716\AA, $\lambda$15748.988\AA, and $\lambda$15765.842\AA), two Al I lines ($\lambda$16718.957\AA, and $\lambda$16750.564\AA), four Si I lines ($\lambda$15888.410\AA, $\lambda$15960.063\AA, $\lambda$16060.009\AA, and $\lambda$16094.787\AA), three Ti I lines ($\lambda$15334.847\AA, $\lambda$15543.756\AA, and $\lambda$15715.573\AA), one Mn I line, $\lambda$15217.0\AA, and 25 Fe I lines.
K and Ca display a small abundance trend as a function of $T_{\rm eff}$ for all lines analyzed, but given that the effects are small, the K I and Ca I lines were kept in the analysis.
In Table \ref{tab:apendix}, we provide a comprehensive list of spectral lines adopted and removed from this work.

\subsection{Abundance uncertainties}  \label{sec:uncertanties}

Uncertainties in the derived abundances can be estimated from the abundance sensitivities to errors in the atmospheric parameters (such as $T_{\rm eff}$, log $g$, and microturbulence), signal-to-noise ratio, and pseudo-continuum normalization \citep{Jofre2019}. We used the abundance sensitivities from Table 4 of \cite{Souto2018} to assess the uncertainties in our derived abundances. 
This error analysis consisted of changing/perturbing the atmospheric parameters based on their typical uncertainties: $\delta$$T_{\rm eff}$ $\pm$ 50~K, $\delta$log $g$ $\pm$0.20 dex, $\delta$[Fe/H] $\pm$0.20 dex, and $\delta$$\xi$ $\pm$0.20 km.s$^{-1}$. We also accounted for uncertainties stemming from changes in the signal-to-noise ratio and pseudo-continuum normalization, using the same methodology as in \cite{Melo2024}. 
To assess the impact of the SNR on our results, we introduced a controlled amount of noise into our synthetic spectra, varying the SNR from 100 to 300 in increments of 20. We then derived the abundances from these spectra to evaluate how each elemental abundance was affected by changes in SNR. For the pseudo-continuum normalization, we adjusted the spectra by $\pm$0.5\% and recalculated the abundances to determine the impact on each element. This analysis was conducted using three representative models for the F-dwarfs ($T_{\rm eff}$ = 6400~K, log $g$ = 4.30), G-dwarfs ($T_{\rm eff}$ = 5777~K, log $g$ = 4.44), and K-dwarfs ($T_{\rm eff}$ = 4800~K, log $g$ = 4.60).

Finally, we combined all uncertainties, from atmospheric parameters, signal-to-noise ratio, and pseudo-continuum normalization, by summing the respective errors in quadrature. The abundance uncertainties for all elements are presented at the bottom of Table \ref{tab:abundances}.
The abundance uncertainty obtained for Fe is 0.06 dex, which is similar to the typical abundance uncertainties for all elements. Potassium abundances have the smallest uncertainties ($\sim$0.04 dex), while the abundances of magnesium and aluminum have the largest uncertainties of 0.10 and 0.08 dex, respectively. 
Another proxy for abundance uncertainties is the standard deviation of the mean. The star-to-star abundance deviation for all elements studied in this work (see Table \ref{tab:abundances}) is about 0.06, in general agreement with the propagated uncertainties obtained.

As discussed in Section \ref{sec:MCMC}, an MCMC analysis of the sample Fe I lines was used as an additional method to determine the stellar metallicities in this study. The MCMC methodology can provide an independent assessment of the uncertainties in the abundance measurements, as it directly estimates the uncertainties in the parameters by means of the standard deviations of the marginalized posteriors. In our case, the mean uncertainties obtained from MCMC for [Fe/H] was $\pm$ 0.06 dex (internal uncertainties + typical standard deviation of the mean), and this is the same as the total uncertainties for iron abundances in Table \ref{tab:abundances}, confirming that the estimated uncertainties using the abundance sensitivities to errors in parameters are probably good estimates of the uncertainties in the derived abundances.  

\subsection{Non-LTE corrections} \label{sec:NLTE}

Departures from local thermodynamic equilibrium (non-LTE) are important to consider and, in particular, significantly impact the optical spectra of evolved metal-poor stars \citep{Asplund2005}. However, in the near-infrared regime, these effects are generally less pronounced for main-sequence FGK dwarfs \citep{Bergemann2012, Osorio2020}. 
The study of \cite{Osorio2020} provides a detailed analysis of multi-element non-LTE calculations for Na, Mg, K, and Ca, crucial for understanding the influence of atomic interactions on stellar abundance determinations in late-type stars. These model grids were essential for developing the synthetic spectra grids utilized in the APOGEE survey's DR17 ASPCAP results. \cite{Osorio2020} discuss that while non-LTE corrections were modest in the H-band ($<$ 0.10 for all studied species), in the optical range, they reached up to 0.7 dex, underscoring the necessity of non-LTE considerations in stellar spectroscopy to achieve accuracies better than 10\%.

We also investigated departures from local thermodynamical equilibrium in our derived abundances for each spectral line studied, applying non-LTE corrections based on \cite{Bergemann2008} and the series of studies by \cite{Bergemann2012, Bergemann2013, Bergemann2015}, which can be accessed from \url{nlte.mpia.de}. 
These corrections utilized 1-D plane-parallel models for representative stars in this study.
We found that non-LTE deviations for Fe, Mg, and Si are negligible and typically smaller than 0.02 dex for each analyzed line. See Table \ref{tab:nlte}.
Manganese showed more significant non-LTE departures across all $T_{\rm eff}$ ranges, varying between +0.06 and +0.16 dex, with a mean correction of +0.12 dex. 
Since the non-LTE corrections are insignificant, we use our derived LTE abundances in all subsequent discussions in this paper, except for the manganese abundances discussed in Section \ref{sec:results}, which were corrected for non-LTE. 

\begin{table}
\caption{Non-LTE abundance corrections.}
\label{tab:nlte}
\begin{tabular}{lrrr}
\hline
Element & $T_{\rm eff}$ = 4800K & $T_{\rm eff}$ = 5777K & $T_{\rm eff}$ = 6400K\\
 & log $g$ = 4.60 & log $g$ = 4.44 & log $g$ = 4.30\\
\hline
Fe & 0.00 & 0.00 & 0.00 \\
Mg & 0.00 & 0.01 & 0.02 \\
Si & 0.00 & 0.00 & -0.01\\
% Ca & 0.00 & 0.04 & 0.08\\
% Ti & 0.06 & 0.10 & 0.12 \\
% Cr & 0.03 & 0.08 & 0.12\\
Mn & 0.06 & 0.12 & 0.16\\
\hline
\end{tabular}
\end{table}

%%%%%%%%%%%%%%%%%%%%%

\onecolumn
{\scriptsize
\renewcommand{\arraystretch}{0.90}
\begin{longtable}{
>{\columncolor[HTML]{FFFFFF}}c 
>{\columncolor[HTML]{FFFFFF}}c 
>{\columncolor[HTML]{FFFFFF}}c 
>{\columncolor[HTML]{FFFFFF}}c 
>{\columncolor[HTML]{FFFFFF}}c 
>{\columncolor[HTML]{FFFFFF}}c 
>{\columncolor[HTML]{FFFFFF}}c 
>{\columncolor[HTML]{FFFFFF}}c 
>{\columncolor[HTML]{FFFFFF}}c 
>{\columncolor[HTML]{FFFFFF}}c 
>{\columncolor[HTML]{FFFFFF}}c 
>{\columncolor[HTML]{FFFFFF}}c 
>{\columncolor[HTML]{FFFFFF}}c 
>{\columncolor[HTML]{FFFFFF}}c }
\caption{Derived Stellar Abundances 
} \\
\hline
2MASS ID & [Fe/H] (MCMC) & [Fe/H] & [C/H] & [Na/H] & [Mg/H] & [Al/H] & [Si/H] & [K/H] & [Ca/H] & [Ti/H] & [V/H] & [Cr/H] & [Mn/H] \\ \hline
2M03450528+2342097 & 0.03 $\pm$ 0.02 & 0.04 & ... & ... & -0.10 & ... & 0.04 & ... & ... & ... & ... & ... & -0.08 \\
2M03445123+2316082 & -0.04 $\pm$ 0.02 & 0.05 & ... & ... & 0.13 & ... & -0.01 & ... & ... & ... & ... & ... & ... \\
2M03435880+2352578 & 0.03 $\pm$ 0.03 & 0.05 & ... & ... & -0.11 & 0.01 & 0.04 & -0.04 & 0.12 & ... & ... & ... & -0.32 \\
2M03501766+2522464 & -0.03 $\pm$ 0.02 & 0.02 & ... & ... & ... & ... & 0.01 & 0.02 & ... & ... & ... & ... & -0.23 \\
2M03475252+2356286 & 0.04 $\pm$ 0.03 & -0.03 & ... & ... & -0.10 & -0.02 & 0.02 & 0.03 & 0.09 & ... & ... & ... & -0.24 \\
2M03482616+2402544 & 0.10 $\pm$ 0.02 & 0.05 & ... & ... & -0.14 & 0.00 & 0.09 & 0.00 & 0.16 & ... & ... & ... & -0.09 \\
2M03444075+2449067 & 0.05 $\pm$ 0.02 & 0.08 & ... & ... & -0.01 & 0.00 & 0.00 & -0.05 & 0.15 & ... & ... & ... & -0.11 \\
2M03394117+2317271 & 0.02 $\pm$ 0.02 & 0.05 & ... & ... & 0.09 & 0.01 & 0.04 & -0.03 & 0.09 & ... & ... & ... & -0.13 \\
2M03463878+2457346 & 0.04 $\pm$ 0.02 & 0.05 & ... & ... & ... & ... & 0.04 & ... & 0.12 & ... & ... & ... & -0.05 \\
2M03491172+2438117 & 0.03 $\pm$ 0.01 & 0.05 & -0.08 & ... & -0.04 & 0.04 & -0.03 & -0.09 & 0.13 & ... & ... & 0.14 & -0.13 \\
2M03504007+2355590 & 0.02 $\pm$ 0.02 & 0.00 & ... & 0.14 & -0.05 & 0.01 & 0.02 & -0.04 & 0.06 & ... & ... & ... & -0.18 \\
2M03385686+2434112 & 0.09 $\pm$ 0.02 & 0.17 & ... & 0.20 & -0.16 & 0.08 & 0.08 & -0.01 & 0.17 & ... & ... & ... & 0.03 \\
2M03462267+2434126 & 0.11 $\pm$ 0.02 & 0.11 & -0.08 & 0.11 & -0.06 & 0.06 & 0.08 & 0.04 & ... & ... & ... & ... & -0.08 \\
2M03445639+2425574 & 0.06 $\pm$ 0.02 & 0.05 & -0.09 & 0.04 & -0.04 & 0.09 & 0.04 & 0.04 & 0.13 & ... & ... & ... & -0.18 \\
2M03462735+2508080 & 0.10 $\pm$ 0.02 & 0.11 & ... & 0.14 & -0.08 & 0.08 & 0.09 & 0.07 & 0.17 & ... & ... & ... & -0.03 \\
2M03483451+2326053 & 0.04 $\pm$ 0.02 & 0.05 & ... & 0.10 & -0.04 & 0.05 & 0.02 & -0.04 & 0.07 & ... & ... & 0.18 & -0.22 \\
2M03481712+2353253 & 0.15 $\pm$ 0.02 & 0.13 & ... & 0.13 & -0.08 & 0.08 & 0.11 & 0.09 & 0.18 & ... & ... & 0.19 & 0.01 \\
2M03464706+2254525 & 0.04 $\pm$ 0.02 & 0.07 & ... & 0.14 & 0.02 & 0.12 & 0.07 & -0.12 & 0.13 & ... & ... & 0.17 & -0.18 \\
2M03465491+2447468 & 0.04 $\pm$ 0.02 & 0.06 & ... & 0.07 & -0.01 & 0.01 & 0.01 & -0.01 & 0.14 & ... & ... & 0.12 & -0.19 \\
2M03465373+2335009 & 0.06 $\pm$ 0.02 & 0.05 & ... & 0.09 & 0.03 & 0.05 & 0.04 & -0.03 & 0.15 & ... & ... & 0.11 & -0.24 \\
2M03441391+2446457 & 0.04 $\pm$ 0.02 & 0.07 & ... & 0.10 & 0.03 & 0.00 & 0.03 & -0.11 & 0.15 & ... & ... & 0.14 & -0.09 \\
2M03474044+2421525 & 0.11 $\pm$ 0.03 & 0.10 & ... & 0.07 & -0.03 & 0.18 & 0.10 & -0.02 & 0.18 & ... & ... & 0.21 & -0.15 \\
2M03450400+2515282 & 0.05 $\pm$ 0.02 & 0.04 & ... & 0.08 & 0.03 & 0.17 & 0.04 & 0.07 & 0.12 & ... & ... & 0.20 & -0.14 \\
2M03440424+2459233 & -0.01 $\pm$ 0.02 & 0.02 & ... & 0.02 & -0.01 & -0.01 & -0.04 & -0.04 & 0.12 & ... & ... & 0.08 & -0.18 \\
2M03440059+2332382 & 0.04 $\pm$ 0.02 & 0.07 & ... & 0.05 & -0.04 & 0.07 & 0.08 & -0.07 & 0.13 & ... & ... & ... & -0.16 \\
2M03493312+2347435 & 0.09 $\pm$ 0.02 & 0.08 & ... & ... & 0.02 & 0.05 & 0.06 & 0.07 & 0.19 & ... & ... & 0.10 & -0.11 \\
2M03433195+2340266 & 0.00 $\pm$ 0.02 & -0.03 & ... & ... & ... & 0.02 & 0.01 & -0.06 & 0.07 & ... & ... & ... & -0.30 \\
2M03405042+2325064 & 0.01 $\pm$ 0.02 & 0.03 & ... & 0.09 & 0.00 & -0.04 & 0.03 & -0.05 & 0.11 & ... & ... & -0.03 & -0.15 \\
2M03433772+2332096 & 0.02 $\pm$ 0.02 & 0.05 & ... & ... & 0.07 & 0.08 & 0.04 & -0.02 & 0.14 & ... & ... & 0.07 & -0.14 \\
2M03481769+2502523 & -0.01 $\pm$ 0.03 & 0.01 & ... & 0.09 & 0.00 & 0.00 & -0.01 & -0.10 & 0.09 & ... & ... & 0.10 & -0.20 \\
2M03403436+2340574 & 0.09 $\pm$ 0.02 & 0.13 & ... & ... & 0.10 & 0.16 & 0.09 & -0.04 & 0.22 & ... & ... & ... & -0.03 \\
2M03474811+2313053 & 0.03 $\pm$ 0.02 & 0.04 & ... & 0.08 & 0.03 & 0.02 & 0.04 & -0.03 & 0.13 & ... & ... & 0.15 & -0.12 \\
2M03502089+2428003 & 0.02 $\pm$ 0.02 & 0.03 & ... & 0.02 & -0.03 & 0.02 & 0.02 & -0.05 & 0.13 & ... & ... & 0.07 & -0.15 \\
2M03454184+2425534 & 0.05 $\pm$ 0.02 & 0.08 & ... & 0.05 & 0.07 & 0.08 & 0.05 & -0.03 & 0.17 & ... & ... & 0.20 & -0.13 \\
2M03502130+2305470 & 0.01 $\pm$ 0.02 & 0.04 & ... & 0.14 & -0.02 & 0.02 & 0.04 & -0.08 & 0.09 & ... & ... & 0.13 & -0.17 \\
2M03444398+2413523 & 0.00 $\pm$ 0.02 & 0.02 & ... & 0.03 & -0.01 & 0.00 & 0.03 & -0.11 & 0.07 & ... & ... & 0.06 & -0.18 \\
2M03392780+2353420 & 0.05 $\pm$ 0.03 & 0.08 & -0.12 & 0.10 & 0.09 & -0.01 & 0.06 & -0.04 & 0.17 & ... & ... & 0.18 & -0.13 \\
2M03433440+2345429 & -0.06 $\pm$ 0.02 & -0.03 & ... & 0.06 & -0.06 & -0.03 & -0.05 & -0.07 & 0.11 & ... & ... & 0.03 & -0.26 \\
2M03435070+2414508 & -0.08 $\pm$ 0.02 & -0.04 & ... & 0.07 & -0.03 & 0.02 & -0.02 & -0.07 & 0.06 & ... & ... & 0.09 & -0.25 \\
2M03470141+2329419 & 0.03 $\pm$ 0.02 & 0.05 & ... & 0.07 & 0.01 & 0.06 & 0.06 & -0.10 & 0.10 & ... & ... & 0.10 & -0.16 \\
2M03492873+2342440 & 0.01 $\pm$ 0.02 & 0.02 & -0.08 & 0.06 & 0.04 & 0.01 & 0.03 & -0.10 & 0.11 & 0.03 & ... & 0.12 & -0.20 \\
2M03495035+2342202 & 0.04 $\pm$ 0.02 & 0.06 & -0.14 & 0.09 & 0.02 & 0.04 & 0.03 & -0.09 & 0.14 & ... & ... & 0.16 & -0.13 \\
2M03505508+2411508 & -0.01 $\pm$ 0.02 & 0.01 & ... & 0.08 & -0.01 & 0.02 & 0.02 & -0.10 & 0.07 & ... & ... & 0.13 & -0.17 \\
2M03532369+2403542 & -0.05 $\pm$ 0.02 & -0.01 & ... & 0.03 & 0.00 & 0.00 & 0.01 & -0.09 & 0.08 & ... & ... & -0.03 & -0.27 \\
2M03490232+2315088 & 0.00 $\pm$ 0.02 & 0.01 & -0.08 & 0.03 & -0.02 & -0.05 & 0.01 & -0.09 & 0.05 & ... & ... & -0.07 & -0.21 \\
2M03461174+2437203 & -0.06 $\pm$ 0.02 & -0.06 & 0.00 & 0.02 & -0.10 & -0.03 & -0.03 & -0.18 & 0.06 & ... & ... & 0.00 & -0.41 \\
2M03444317+2552319 & 0.00 $\pm$ 0.02 & 0.02 & ... & 0.03 & 0.02 & 0.01 & 0.00 & -0.15 & 0.10 & 0.06 & ... & 0.06 & -0.22 \\
2M03440484+2416318 & -0.01 $\pm$ 0.02 & 0.00 & ... & 0.10 & 0.03 & 0.03 & -0.02 & -0.06 & 0.09 & ... & ... & ... & -0.27 \\
2M03573331+2403114 & 0.00 $\pm$ 0.02 & 0.00 & 0.02 & 0.04 & 0.02 & -0.03 & 0.01 & -0.17 & 0.05 & 0.06 & ... & 0.02 & -0.20 \\
2M03430293+2440110 & 0.04 $\pm$ 0.02 & 0.03 & -0.04 & 0.04 & 0.06 & 0.02 & 0.06 & -0.08 & 0.10 & ... & ... & 0.08 & -0.18 \\
2M03450326+2350219 & 0.02 $\pm$ 0.02 & 0.03 & ... & 0.02 & -0.04 & -0.04 & 0.04 & -0.09 & 0.10 & ... & ... & 0.07 & -0.17 \\
2M03513903+2245010 & 0.00 $\pm$ 0.02 & 0.00 & 0.05 & 0.04 & 0.02 & 0.02 & 0.05 & -0.14 & 0.05 & 0.07 & ... & 0.12 & -0.21 \\
2M03363030+2400440 & 0.02 $\pm$ 0.02 & 0.03 & ... & 0.12 & 0.02 & 0.00 & 0.04 & -0.17 & 0.12 & 0.01 & ... & 0.08 & -0.23 \\
2M03403072+2429143 & -0.10 $\pm$ 0.02 & -0.08 & ... & 0.02 & -0.16 & ... & -0.01 & -0.20 & 0.02 & ... & ... & ... & -0.36 \\
2M03444394+2529574 & -0.02 $\pm$ 0.03 & -0.01 & -0.12 & 0.02 & -0.08 & -0.03 & 0.02 & -0.14 & 0.10 & ... & ... & -0.07 & -0.27 \\
2M03470678+2342546 & 0.05 $\pm$ 0.02 & 0.03 & ... & 0.04 & 0.12 & -0.04 & 0.01 & -0.11 & 0.06 & 0.04 & ... & 0.07 & -0.21 \\
2M03471480+2522186 & -0.06 $\pm$ 0.02 & -0.06 & 0.01 & 0.12 & -0.06 & ... & -0.01 & -0.20 & 0.01 & ... & ... & ... & -0.38 \\
2M03434901+2543466 & 0.00 $\pm$ 0.02 & -0.01 & ... & 0.03 & -0.01 & -0.04 & -0.03 & -0.20 & 0.07 & 0.06 & 0.08 & 0.01 & -0.28 \\
2M03511685+2349357 & 0.05 $\pm$ 0.02 & 0.06 & -0.07 & ... & 0.11 & -0.02 & 0.07 & -0.14 & 0.08 & ... & ... & 0.06 & -0.19 \\
2M03441120+2322455 & 0.08 $\pm$ 0.02 & 0.00 & ... & 0.06 & -0.12 & 0.00 & 0.04 & -0.18 & 0.10 & ... & ... & -0.11 & -0.34 \\
2M03505143+2319447 & 0.05 $\pm$ 0.02 & 0.05 & ... & 0.05 & 0.04 & 0.00 & 0.07 & -0.12 & 0.10 & -0.02 & ... & 0.16 & -0.14 \\
2M03452219+2328182 & 0.04 $\pm$ 0.02 & 0.04 & ... & ... & 0.03 & -0.02 & 0.04 & -0.09 & 0.02 & -0.05 & ... & ... & -0.20 \\
2M03440509+2529017 & 0.01 $\pm$ 0.03 & -0.02 & -0.13 & 0.01 & 0.00 & -0.07 & 0.00 & -0.17 & 0.04 & ... & 0.12 & 0.02 & -0.23 \\
2M03404256+2542197 & 0.00 $\pm$ 0.03 & -0.02 & -0.09 & 0.06 & -0.03 & -0.06 & -0.01 & -0.19 & 0.03 & -0.05 & ... & 0.02 & -0.27 \\
2M03471352+2342515 & 0.04 $\pm$ 0.02 & 0.03 & ... & 0.05 & 0.04 & -0.02 & 0.00 & -0.16 & 0.09 & 0.00 & 0.12 & 0.05 & -0.12 \\
2M03463938+2401468 & 0.04 $\pm$ 0.03 & 0.06 & -0.03 & 0.09 & 0.02 & -0.06 & 0.02 & -0.03 & 0.09 & 0.03 & 0.01 & 0.11 & ... \\
2M03420470+2553091 & 0.07 $\pm$ 0.02 & 0.05 & ... & 0.03 & 0.05 & 0.01 & 0.01 & -0.12 & 0.11 & -0.04 & -0.10 & 0.13 & -0.16 \\
2M03555603+2334021 & 0.04 $\pm$ 0.02 & 0.05 & -0.07 & ... & 0.09 & -0.08 & 0.04 & -0.19 & 0.11 & 0.01 & 0.12 & 0.07 & -0.17 \\
2M03415906+2555153 & 0.05 $\pm$ 0.02 & 0.04 & ... & ... & 0.08 & -0.06 & 0.05 & -0.17 & 0.10 & 0.00 & -0.10 & 0.10 & -0.17 \\
2M03422759+2502492 & 0.05 $\pm$ 0.03 & 0.05 & ... & 0.06 & 0.06 & 0.00 & 0.05 & -0.10 & 0.12 & -0.03 & ... & 0.04 & -0.15 \\
2M03460649+2434027 & 0.06 $\pm$ 0.03 & 0.05 & -0.05 & 0.05 & 0.02 & -0.02 & 0.05 & -0.19 & 0.10 & 0.05 & 0.08 & 0.10 & -0.16 \\
2M03503457+2430281 & 0.02 $\pm$ 0.03 & 0.02 & ... & 0.07 & 0.01 & -0.04 & 0.03 & -0.12 & 0.04 & -0.03 & ... & -0.02 & -0.19 \\
2M03401202+2538321 & -0.10 $\pm$ 0.03 & -0.08 & 0.02 & 0.10 & -0.09 & ... & -0.03 & ... & ... & ... & 0.08 & ... & ... \\
2M03513927+2432561 & 0.03 $\pm$ 0.03 & 0.02 & ... & 0.04 & 0.01 & -0.08 & 0.06 & -0.19 & 0.01 & -0.07 & ... & 0.09 & -0.23 \\
2M03540892+2420011 & -0.01 $\pm$ 0.03 & 0.00 & -0.04 & 0.07 & -0.02 & ... & 0.03 & -0.23 & 0.02 & -0.16 & ... & -0.04 & -0.20 \\
2M03413245+2309422 & 0.02 $\pm$ 0.01 & 0.02 & ... & 0.06 & 0.00 & -0.07 & -0.02 & -0.16 & 0.03 & -0.09 & ... & -0.10 & ... \\
2M03452957+2345379 & 0.05 $\pm$ 0.03 & 0.05 & ... & 0.03 & 0.07 & -0.11 & 0.05 & -0.15 & 0.09 & -0.02 & ... & 0.16 & -0.13 \\
2M03405126+2335543 & 0.08 $\pm$ 0.03 & 0.08 & ... & ... & 0.13 & -0.01 & 0.02 & -0.15 & 0.08 & 0.00 & 0.10 & 0.13 & -0.16 \\
2M03432662+2459395 & 0.01 $\pm$ 0.04 & 0.03 & ... & 0.08 & 0.02 & -0.05 & 0.02 & -0.09 & 0.06 & -0.08 & -0.10 & 0.05 & ... \\
2M03455048+2352262 & 0.03 $\pm$ 0.03 & 0.04 & -0.16 & ... & -0.04 & -0.02 & 0.03 & -0.23 & 0.04 & -0.04 & -0.04 & 0.09 & -0.19 \\
\hline
Mean Cluster values & 0.03 & 0.03 & -0.06 & 0.07 & -0.01 & 0.00 & 0.03 & -0.09 & 0.10 & -0.01 & 0.03 & 0.08 & -0.18 \\
Median Cluster values & 0.03 & 0.04 & -0.07 & 0.06 & 0.00 & 0.00 & 0.03 & -0.09 & 0.10 & 0.00 & 0.08 & 0.09 & -0.18 \\
Median Absolute Deviation (MAD) & 0.02 & 0.02 &  0.08 & 0.04 & 0.04 & 0.02 & 0.02 & 0.06 & 0.03 &  0.00 & 0.08 & 0.07 & 0.05 \\
Standard deviation of the mean & 0.05 & 0.04 & 0.06 & 0.04 & 0.07 & 0.07 & 0.03 & 0.08 & 0.05 & 0.06 & 0.09 & 0.08 & 0.08 \\
Propagated Uncertainties (mean values) & 0.06 & 0.06 & 0.03 & 0.07 & 0.10 & 0.08 & 0.07 & 0.03 & 0.05 & 0.06 & 0.06 & 0.05 & 0.04 \\
\hline
\label{tab:abundances}
\end{longtable}
}
\twocolumn

\section{Results}  \label{sec:results}

In this section, we present abundance results for twelve elements (C, Na, Mg, Al, Si, K, Ca, Ti, V, Cr, Mn, and Fe) in 80 F, G, and K dwarf stars from the Pleiades open cluster. The individual stellar abundances are presented in Table \ref{tab:abundances}, where the mean abundances, standard deviations, and uncertainties obtained for the cluster are presented in the last lines of Table \ref{tab:abundances}.

We derived metallicities using two methods: BACCHUS in semi-automatic mode and an MCMC analysis. Through our abundance analysis using BACCHUS, we derived a mean iron abundance (a proxy for metallicity) for the cluster of $<$[Fe/H]$>$ = 0.03 $\pm$ 0.04 dex, indicating that the Pleiades open cluster is slightly metal-rich. Using a two-dimensional MCMC analysis that considers both [Fe/H] and vsin($i$), we obtained the same mean metallicity result with a slightly higher standard deviation: $<$[Fe/H]$>$ = 0.03 $\pm$ 0.05 dex. 
In Figure \ref{fig:vsinia}, the upper panel provides a direct comparison of our BACCHUS derived [Fe/H] (x-axis) with that obtained from MCMC (y-axis). The color bar denotes $T_{\rm eff}$, while the lower panel illustrates the residual difference between [Fe/H] from semi-automatic BACCHUS analysis and [Fe/H] from MCMC. Our results demonstrate excellent agreement in the metallicity scale, with [Fe/H](BACCHUS - MCMC) = 0.01 $\pm$ 0.02 dex, and we see no systematic trends with $T_{\rm eff}$.
In conclusion, the derived metallicities are consistent across methods, and these results align well with previous measurements from the literature (\citealt{Soderblom2009, Takeda2016, Spina2018}).

For the overall metal content of the cluster, denoted as [M/H], we calculate a value of 0.01 $\pm$ 0.06, reinforcing iron as a reliable proxy for the cluster metallicity. The average alpha-element to iron  abundance ratio, calculated as the mean of [Mg/Fe], [Si/Fe], and [Ca/Fe], is [$\alpha$/Fe] = 0.01 $\pm$ 0.05. For odd-z elements (Na, Al, K), we find [odd-z/Fe] = -0.04 $\pm$ 0.08, and for the iron-peak elements V, Cr, and Mn, [iron peak/Fe] = -0.02 $\pm$ 0.08. These values, indicative of yields from both Type I and II supernovae, approach solar levels—an unsurprising result, given the Pleiades' proximity to the Sun (135 pc; \citealt{Lodieu2019}) and its relative youth (112 Myr; \citealt{Dahm2015}). We reference solar abundance values from \cite{Asplund2021}.

Examining the individual elements, we find their abundances to be in close proximity to solar values, such as [Ti/H] = -0.01 $\pm$ 0.06, [Mg/H] = 0.00 $\pm$ 0.07, [Al/H] = 0.00 $\pm$ 0.07, [Si/H] = 0.03 $\pm$ 0.03, and [V/H] = 0.03 $\pm$ 0.09. Some elements exhibit slightly enhanced abundances, including [Na/H] = 0.07 $\pm$ 0.04, [Cr/H] = 0.08 $\pm$ 0.08, and [Ca/H] = 0.10 $\pm$ 0.05 dex. Others show slightly depleted abundance ratios, such as [C/H] = -0.06 $\pm$ 0.06, [Mn/H] = -0.08 $\pm$ 0.08 dex, and [K/H] = -0.09 $\pm$ 0.08. 
Table \ref{tab:abundances} lists the stellar abundances determined in this study. 

\subsection{Comparisons with the literature}   \label{subsec:com-literature}
\subsubsection{APOGEE DR17}  \label{sec:comp_DR17}

In this section, we compare our results with those from SDSS-IV APOGEE DR17.
While APOGEE primarily targeted red giants, it also included numerous main-sequence stars.
Although the SDSS-IV has ended, DR17 results are used by the community,  and the ASPCAP pipeline continues to be used as one of the pipelines for parameters and abundance determinations in the ongoing SDSS-V Milky Way Mapper (MWM) survey, that now has an important component focusing on main-sequence stars. 
Consequently, our results can provide valuable calibration data for the abundances derived by their pipelines.

The comparison of our computed vsin($i$) using MCMC (x-axis) and those from APOGEE DR17 (y-axis) is shown in the lower panel of Figure \ref{fig:vsinia}. The results generally agree, with minor scatter, vsin($i$) (this work - APOGEE DR17) = 0.45 $\pm$ 0.94 km.s$^{-1}$. There is a noticeable trend in vsin($i$) as a function of $T_{\rm eff}$, where warmer F-dwarfs exhibit higher vsin($i$) ($\gtrsim$ 20 km.s$^{-1}$)) compared to cooler K-dwarfs, which show lower values ($\lesssim$ 10 km.s$^{-1}$).

\begin{figure*}
  \centering
{\includegraphics[width=0.31\textwidth]{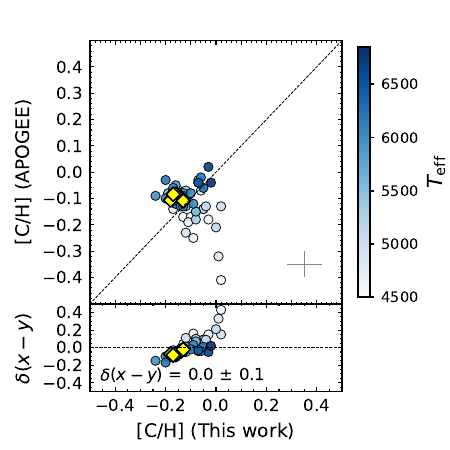}\label{fig:delta_a}} 
{\includegraphics[width=0.31\textwidth]{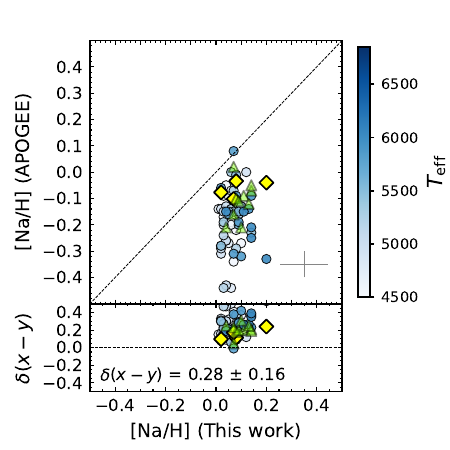}\label{fig:delta_b}}
{\includegraphics[width=0.31\textwidth]{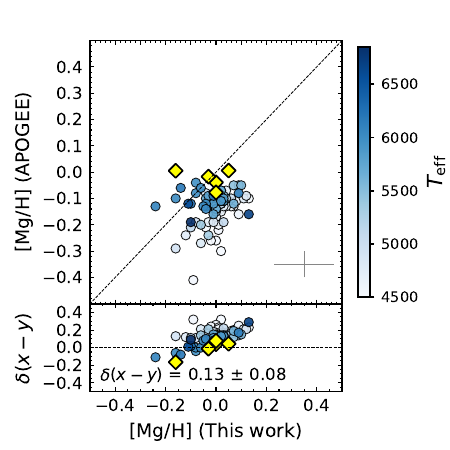}\label{fig:delta_c}}\\
{\includegraphics[width=0.31\textwidth]{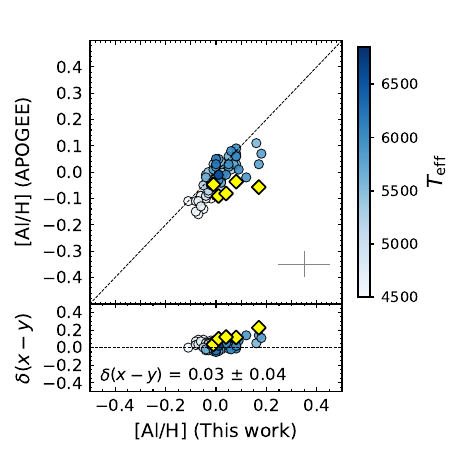}\label{fig:delta_d}}
{\includegraphics[width=0.31\textwidth]{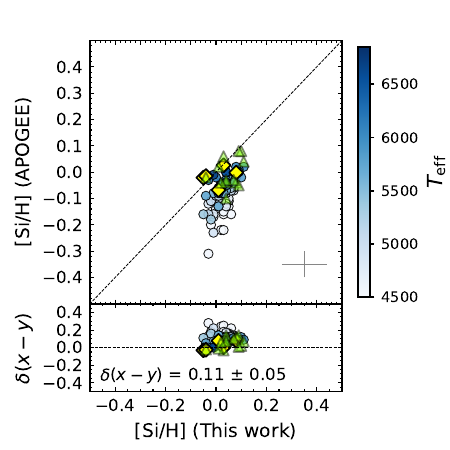}\label{fig:delta_e}}
{\includegraphics[width=0.31\textwidth]{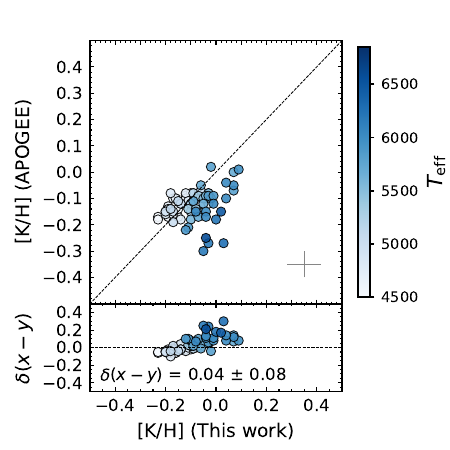}\label{fig:delta_f}}\\
{\includegraphics[width=0.31\textwidth]{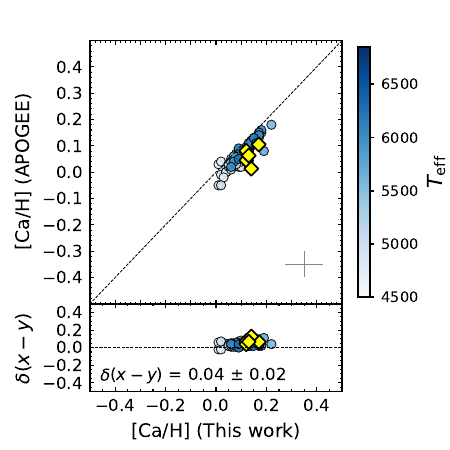}\label{fig:delta_g}} 
{\includegraphics[width=0.31\textwidth]{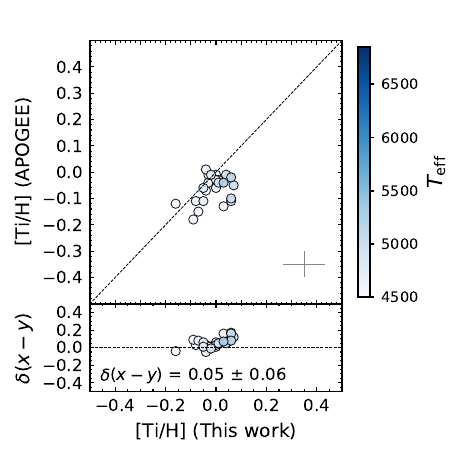}\label{fig:delta_h}}
{\includegraphics[width=0.31\textwidth]{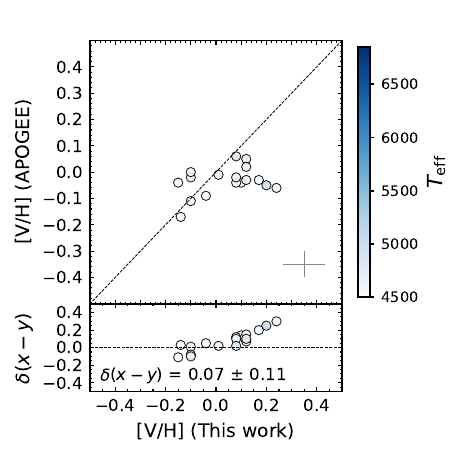}\label{fig:delta_i}}\\
{\includegraphics[width=0.31\textwidth]{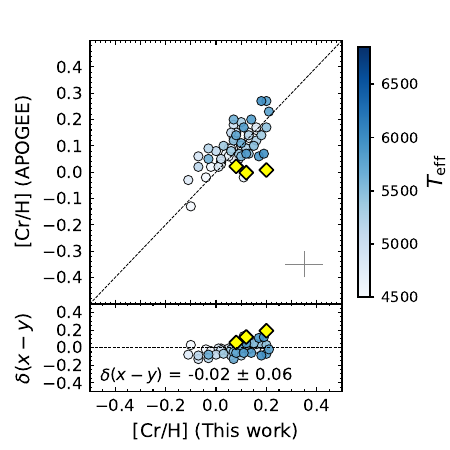}\label{fig:delta_j}}
{\includegraphics[width=0.31\textwidth]{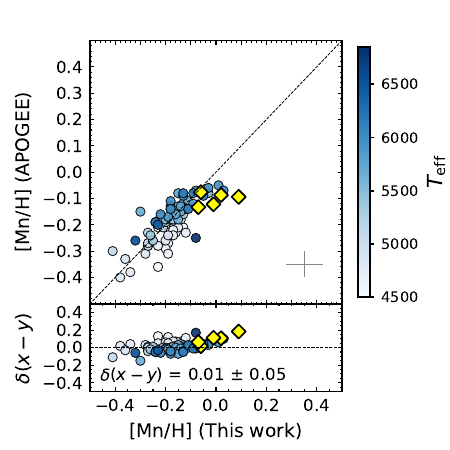}\label{fig:delta_k}}
{\includegraphics[width=0.31\textwidth]{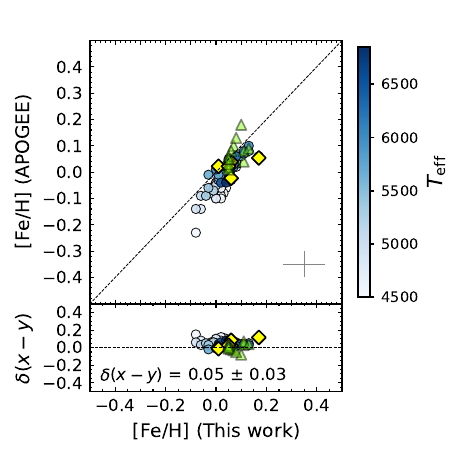}\label{fig:delta_l}}
\caption{Abundance comparisons of this work and those from the literature (APOGEE DR17, circles with color bar; Spina \textit{et al.} (2018) as a yellow diamond; Soderblom \textit{et al.} (2009) as a green triangle).  Each panel represents an individual element. Same format as Figure \ref{fig:vsinia}.}
\label{fig:dr17_comp}
\end{figure*}

Figure \ref{fig:dr17_comp} presents a one-to-one abundance diagram, plotting our results on the x-axis versus APOGEE DR17 (uncalibrated), \cite{Soderblom2009}, and \cite{Spina2018}  results on the y-axis. (This section focuses on the comparisons with DR17, while Section \ref{sec:comp_Optical} will compare with results from the optical.) 
Each subplot's bottom panel shows the residual difference ($\delta$) between our work and APOGEE DR17, with a color bar indicating $T_{\rm eff}$. The comparison of the abundance results generally shows good agreement, particularly for Ca ($\delta$ (this work - literature) = 0.04 $\pm$ 0.02), Cr ($\delta$ = -0.02 $\pm$ 0.06), Mn ($\delta$ = 0.01 $\pm$ 0.05), and Fe ($\delta$ = 0.05 $\pm$ 0.03).
However, some elements like C ($\delta$ = 0.00 $\pm$ 0.10), K ($\delta$ = 0.04 $\pm$ 0.08) and Mg ($\delta$ = 0.13 $\pm$ 0.08) exhibit more scatter, and Na ($\delta$ = 0.28 $\pm$ 0.16) shows significant difference. 
However, it is clear from the figure that the DR17 results for C, Na, Mg, Si, and K show a systematic decrease in the abundances, which is more pronounced for stars with lower $T_{\rm eff}$, as seen from the light-colored circles, while this work results (x-axis) are clumped around 0.00. 
This suggests that DR17 may exhibit abundance trends influenced by effective temperature, as discussed in Section \ref{sec:trends}.

It is notable that the Na abundances for the studied Pleiades stars in DR17 have very large offsets and systematic trends; Na abundances are derived from two weak lines in the APOGEE spectra. These systematic trends may be attributed to the challenges ASPCAP faces in accurately measuring weak lines. Such trends result in mean Na abundances with very large scatter, which is not expected for stars in an open cluster (standard deviation of the mean of 0.23). In contrast, in this work, we obtain a much more reasonable standard deviation of the mean A(Na) of 0.04 for the studied Pleiades stars.

%--------------------------------------------
%--------------------------------------------

\subsubsection{Optical high-resolution abundance studies}  \label{sec:comp_Optical}

The Pleiades is one of the most studied open clusters in our Galaxy. 
Despite this, there remains a lack of comprehensive abundance studies spanning the cluster’s extensive range of stellar masses. Previous studies have primarily focused on the cluster's luminous peculiar stars. Notably, \cite{Hui-Bon-Hoa1998} and \cite{Gebran2008} have conducted abundance analyses on A-type stars, both normal and chemically peculiar (Am), as well as F-type stars using high-resolution optical spectroscopy. These studies determined abundances for a variety of elements, including C, N, O, Na, Mg, Si, Ca, Sc, Ti, V, Cr, Mn, Fe, Co, Ni, Sr, Y, Zr, and Ba. Additionally, \cite{Soderblom2009} and \cite{Takeda2016} have characterized the abundances of C, O, Na, Fe, Si, Ni, and Ti in FGK dwarf stars, also employing high-resolution optical spectra.

More recently, \cite{Spina2018} explored chemical inhomogeneities in the Pleiades, potentially caused by planetary engulfment. The authors utilized high-resolution spectroscopy from the UVES at the Very Large Telescope of the European Southern Observatory. 
The five stars studied by \cite{Spina2018} are also examined in this work: 2M03385686+2434112, 2M03440424+2459233, 2M03450400+2515282, 2M03465491+2447468, and 2M03491172+2438117. This work's atmospheric parameters show good agreement with \cite{Spina2018}, where $\delta$$T_{\rm eff}$ = -31 $\pm$ 39K, and $\delta$log $g$ = -0.06 $\pm$ 0.04 dex ($\delta$ represents the mean difference, This Work - \citealt{Spina2018}).
Overall, our abundance results agree well with their results, as shown by the yellow diamond symbols  in Figure \ref{fig:dr17_comp}. 
Elements exhibiting significant disparities in this comparison are Na ($\delta$ = 0.16 $\pm$ 0.06), Al ($\delta$ = 0.12 $\pm$ 0.06), Cr ($\delta$ = 0.12 $\pm$ 0.05), and Mn ($\delta$ = 0.10 $\pm$ 0.06).  The presence of weak lines in the APOGEE spectra for Na and Cr within the atmospheric parameter range of our studied sample may indicate additional challenges in measuring their abundances. Elements like C, Mg, Si, and Fe exhibit mean abundance differences smaller than 0.06 dex, indicating a good agreement within the uncertainties.

Thirteen F and G dwarfs in this work were previously studied in \cite{Soderblom2009}. (They are marked on Table \ref{tab:sample}). 
The latter study determined abundances of Fe, Si, Ni, Ti, and Na from the Hamilton optical echelle high-resolution spectrograph (R=40,000) in 20 Pleiades solar-like stars having low vsin($i$) values. 
Comparisons between our results and those from \cite{Soderblom2009} generally show good agreement, with $\delta$[Fe/H] = 0.00 $\pm$ 0.04 and $\delta$[Si/H] = 0.06 $\pm$ 0.06. However, a notable discrepancy exists for Na, where $\delta$[Na/H] = 0.21 $\pm$ 0.06, indicating significantly lower values in \cite{Soderblom2009}, as seen in Figure as a green triangle in Figure \ref{fig:dr17_comp}.
The observed offset in Na abundances between this study and those reported by \cite{Soderblom2009} and \cite{Spina2018} could be attributed to non-LTE effects impacting the optical Na I lines used in their analyses, as these may be significantly affected by departures from LTE \citep{Asplund2005}.

%--------------------------------------------
\begin{figure*}
    \includegraphics[width=0.6\linewidth]{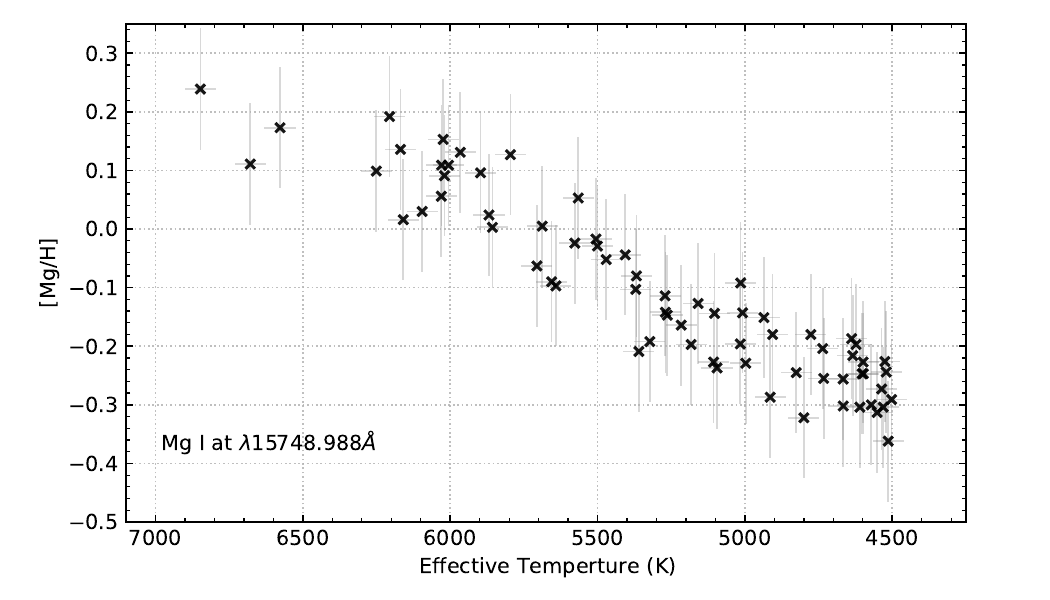}
    \includegraphics[width=0.38\linewidth]{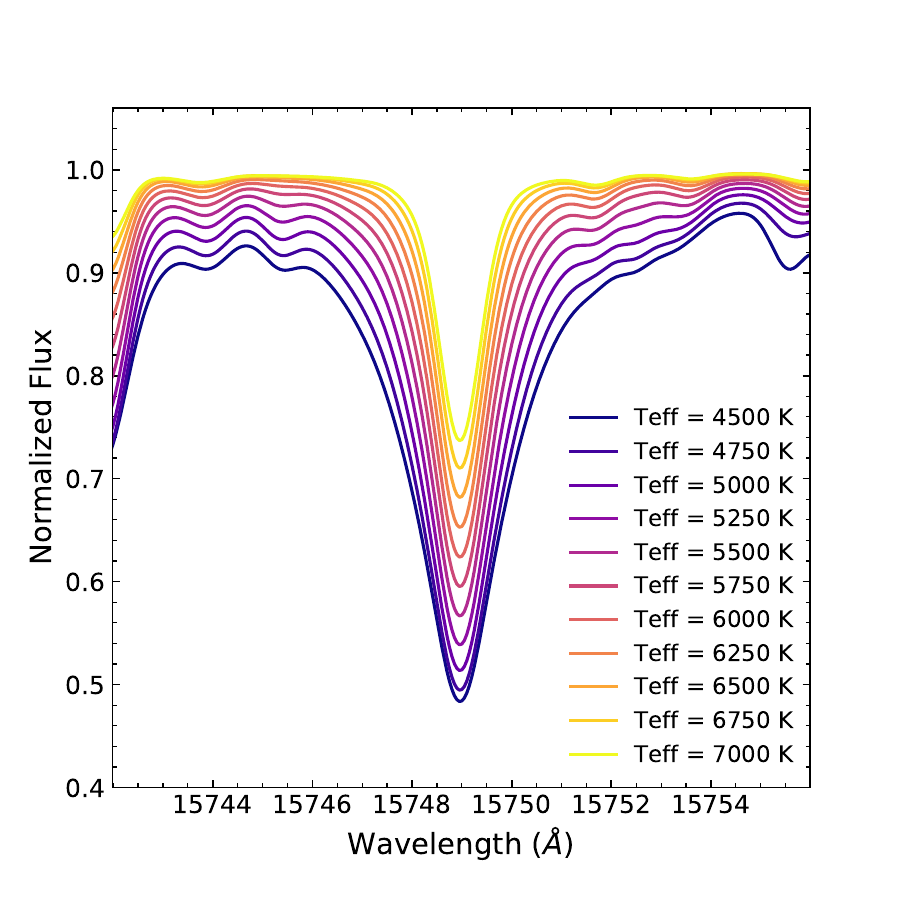}
    \includegraphics[width=0.6\linewidth]{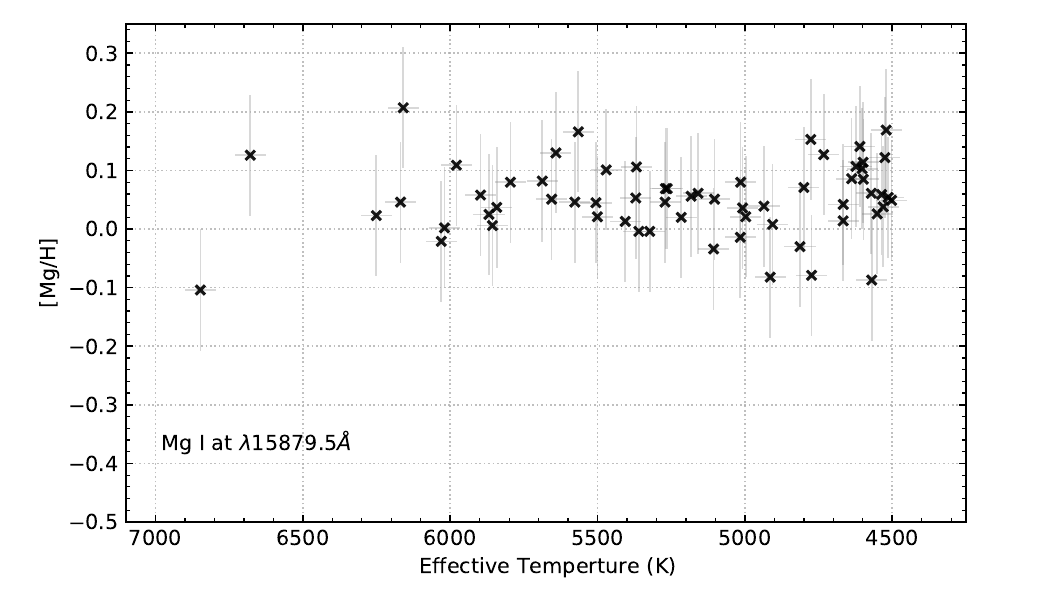}
    \includegraphics[width=0.38\linewidth]{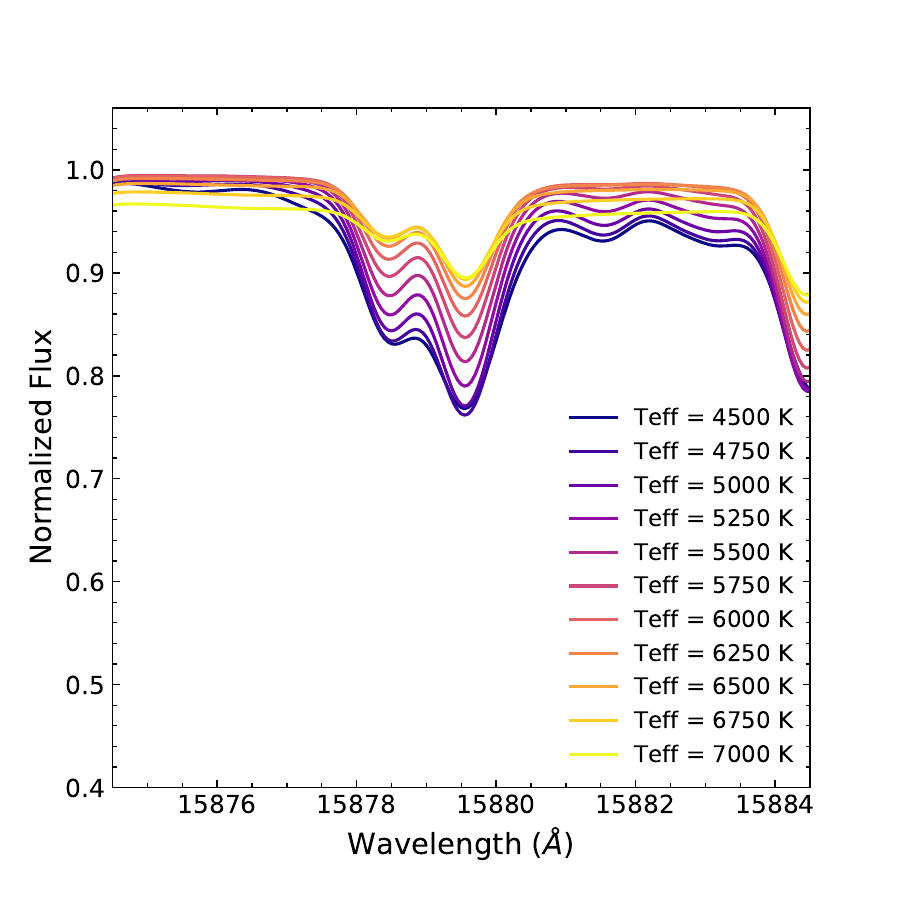}
  \caption{Left panel: Values of [Mg/H] $vs.$ $T_{\rm eff}$, with the left upper panel showing abundances derived from the Mg I line at $\lambda$15748.988\AA{}, while the left lower panel shows results for the  $\lambda$15879.5\AA{} line. Right panel: Spectral syntheses for different values of $T_{\rm eff}$, ranging from 4500 to 7000 K in steps of 250 K, assuming fixed values for log $g$ = 4.50, $\xi$ = 1.0 km.s$^{-1}$, and [Fe/H] = 0.00.  The right upper panel shows the Mg I line at $\lambda$15748.988\AA{}, and the right lower panel shows the Mg I line at $\lambda$15879.5\AA{}.} 
\label{fig:trend_Mg}
\end{figure*}
%--------------------------------------------

%--------------------------------------------
\begin{figure*}
  \centering
  \subfloat{\includegraphics[width=0.95\textwidth]{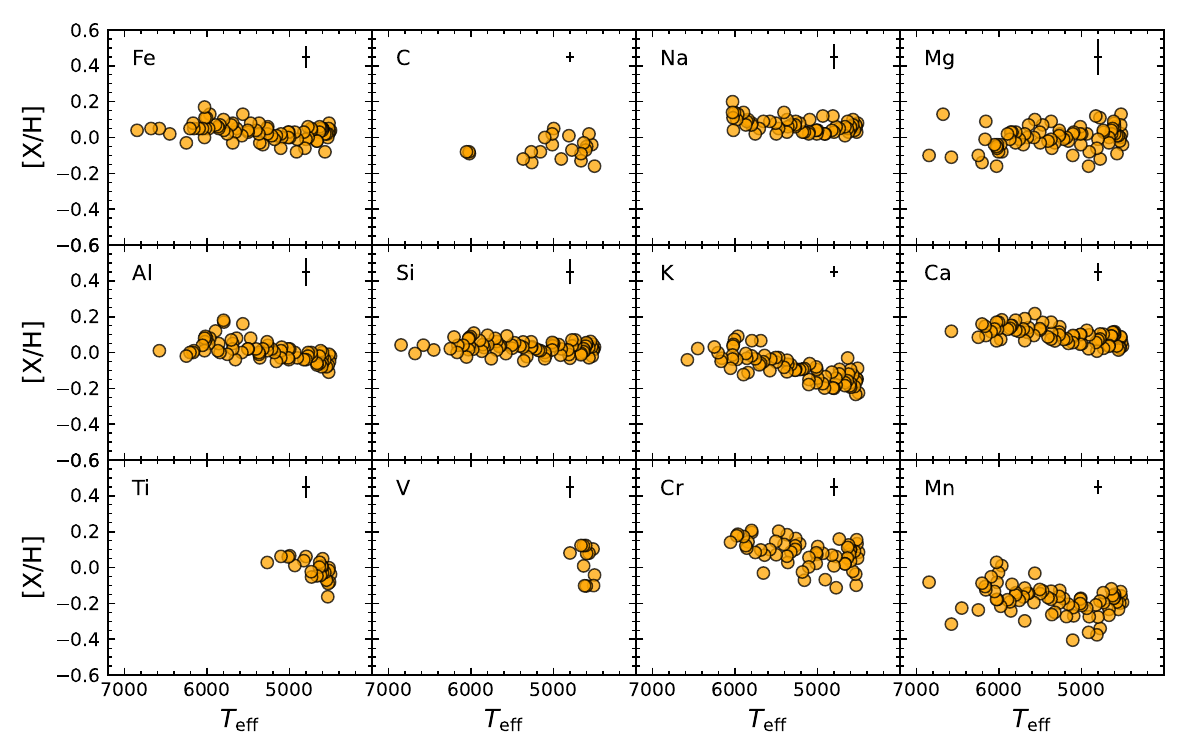}}\label{fig:X_teff} \\
  \caption{Derived abundances ([X/H]) of the studied stars as a function of effective temperature.}
  \label{fig:teffxelements}
\end{figure*}
%--------------------------------------------

%--------------------------------------------
\begin{figure}
    \includegraphics[width=0.95\linewidth]{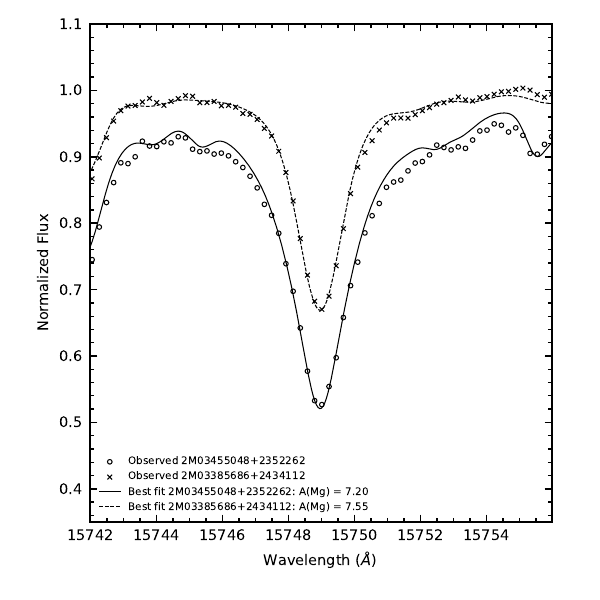}
  \caption{A small portion of two APOGEE spectra displaying a single spectral line with the observed and best-fit synthetic spectra for an F-dwarf (2M03385686+2434112; crosses show the observed spectrum and the dashed black line the best-fit spectrum) and a K-dwarf (2M03455048+2352262; circles represent the observed data and the solid black line is the best-fit synthetic spectrum), centered on the Mg I line at $\lambda$15748.988\AA{}.}
\label{fig:lines_comparison}
\end{figure}
%--------------------------------------------

\section{Discussion} \label{sec:discussion}

\subsection{Abundance trends as a function of $T_{\rm eff}$}  \label{sec:trends}

%Open clusters serve as essential astrophysical laboratories for examining physical processes during stellar evolution. They provide unique opportunities to validate methodologies for determining stellar abundances across stars with varied masses, effective temperatures, and surface gravities. Analyzing how stellar abundances correlate with effective temperatures, surface gravities, mass, and evolutionary stages can reveal systematic errors in abundance determinations or signatures of abundance variations. Notably, variations in [X/Fe] as a function of either $T_{\rm eff}$, log $g$, or stellar mass offer insights into potential mechanisms such as atomic diffusion (\citealt{Martin2017,Souto2018,Gao2018,Semenova2020,Souto2021}) or dredge-up processes (\citealt{Souto2019,Loaiza-Tacuri2023}). 

In Figure \ref{fig:trend_Mg} left panels, we illustrate how magnesium abundances change with $T_{\rm eff}$ for two specific Mg I lines: the stronger line $\lambda$15748.988 \AA{} (top panel) and the weaker line $\lambda$15879.5 \AA{} (bottom panel). 
For the stronger line, we see that 
as $T_{\rm eff}$ decreases, the Mg abundance derived from the line $\lambda$15748.988\AA{} declines strongly. For $T_{\rm eff}$ values exceeding 6000K, the Mg abundances do not follow the same trend, becoming approximately constant.
In contrast, the bottom-left panel shows the behavior of the Mg abundances derived from the weaker Mg I line at $\lambda$15879.5\AA{}. Here we see no significant trend of the Mg abundances as a function of $T_{\rm eff}$.

On the right panel of Figure \ref{fig:trend_Mg}, we display the synthetic line profiles of these Mg I lines for $T_{\rm effs}$ ranging from 4500 to 7000~K in steps of 250~K, while fixed values for log $g$ = 4.50, $\xi$ = 1.0 km.s$^{-1}$, and [Fe/H] = 0.00 are adopted. The yellow-ish solid lines indicate warmer, and the purple-ish solid lines represent cooler stars with their respective spectral synthesis. 
For the top-right panel, we see the Mg I line profiles getting broader at the wings and having a significant pseudo-continuum suppression of each line as $T_{\rm eff}$ decreases. As a consequence, the FWHM for K dwarfs is about twice that of those late F dwarfs. 
The observed line broadening profile is primarily influenced by the rise in atomic collisions in the stellar atmosphere as stars become more compact along the cooler main sequence. In the bottom-right panel, we can see that the line broadening and pseudo-continuum depletion for the weaker Mg I at $\lambda$15879.5\AA{} are considerably smaller compared to the top-right panel.

We find a similar trend for the Mg abundances derived from the Mg I lines at $\lambda$15740.716 \AA{}, and $\lambda$15765.84 \AA{} similarly to  Figure \ref{fig:trend_Mg} top left panel. 
However, the Mg I lines at $\lambda$15886.2\AA{} and $\lambda$15954.477\AA{} do not exhibit a substantial trend with $T_{\rm eff}$ and resemble the behavior seen in the bottom-left panel of Figure \ref{fig:trend_Mg}.
Given that the Mg abundance of stars in an open cluster are not expected to vary, it appears that a physical or analytical factor strongly influences the profiles of the Mg I lines at $\lambda$15740.176\AA{}, $\lambda$15748.988\AA{}, and $\lambda$15765.842\AA{}, and we are compensating this effect by changing the Mg abundances to roughly fit the lines.
In Appendix \ref{tab:apendix}, we list all spectral lines adopted in this study that exhibit no or weak abundance trends with $T_{\rm eff}$, as well as those lines excluded due to strong trends with $T_{\rm eff}$.

A comparison of the relative line strengths between the included and excluded lines in Appendix \ref{tab:apendix}, within each element, reveals that the excluded lines are the stronger lines (the division is quite striking for C I, Mg I, Al I, Si I, and Ti I).  As the stronger lines will form higher in the stellar atmospheres, the trends with $T_{\rm eff}$ may be related to the depth-of-formation and perhaps involve non-LTE (although in our case, departures from LTE are small) or, for example, be related to inadequacies in our treatment of line broadening. 

As discussed above, in this study, we only use selected lines with small or no significant abundance trends with the effective temperature (see Appendix \ref{tab:apendix}). Figure \ref{fig:teffxelements} provides a visual overview of all of the elemental abundances determined here as a function of effective temperature.  Beginning with Fe, whose abundance is derived from the largest number of lines (31), no trend is found within a small abundance scatter of $\sim\pm$0.05 dex.  Carbon abundances rely on two rather weak C I lines that are measured here over a somewhat restricted $T_{\rm eff}$ range and display no measurable trend with temperature.  The odd-Z elements Na and Al are represented by only 2 lines and 1 line, respectively, with the Na I lines being quite weak, while the Al I line is stronger and well-defined.  Both sets of abundances exhibit insignificant slopes (within our measurement uncertainties).  We can see that the selected Mg I lines in this study (as discussed above) resulted in Mg abundances without significant trends with temperature, while
the cluster Si abundances, in particular, are characterized by a small scatter of $\pm$0.03 dex.  Titanium and vanadium are both measured over a narrow range in effective temperature such that any inherent dependencies on $T_{\rm eff}$ are not detected.  The remaining four elements, K, Ca, Cr, and Mn, all display only small $T_{\rm eff}$ trends that result in increasing abundances with increasing temperature and all having slopes of $\sim$+0.1 dex or less per 1000 K.    
Chromium abundances based upon the APOGEE spectra can be, in general, uncertain due to the fact that Cr abundances are based on one weak Cr I line, which can be impacted by any nearby weak unidentified features.  The other elemental abundances that decrease slightly with decreasing temperature (K, Ca, and Mn) are derived from well-defined lines with weak to moderate line strengths. These results highlight the usefulness of open clusters as checks on the spectroscopic analyses of dwarf stars using the APOGEE spectra.

\subsection{\textbf{Investigating potential causes of abundance trends}}  \label{sec:trends_causes}
\subsubsection*{\textbf{Line profile fitting}}

To investigate possible systematic errors in the best-fit syntheses in our abundance measurements across different stellar classes, we compared the 
fits obtained for a warmer F- and a cooler K-dwarf. This is illustrated in Figure \ref{fig:lines_comparison}, which displays our synthetic best fits for the Mg I line at $\lambda$15748.988 \AA{} in two stars with different effective temperatures: 
2M03385686+2434112 ($T_{\rm eff}$ = 6029~K) and 2M03455048+2352262 ($T_{\rm eff}$ = 4502~K).  The APOGEE spectra of both stars have very high SNR, 530 and 217, respectively. 
The crosses and dashed black line represent the observed and best-fit spectra for 2M03385686+2434112, respectively, while the open  circles and solid black line represent the observed and best-fit spectra for 2M03455048+2352262.
Our syntheses reproduce the Mg I lines with fidelity. For example, the chi-square obtained for 2M03385686+2434112 is 14.0, and for 2M03455048+2352262, it is 5.8, which is quite similar, even though the line intensity changes for both stars. 
As illustrated in the right panels of Figure \ref{fig:trend_Mg}, we also see that
the wings of the cooler main sequence stars are broader, where the equivalent width (EW) is 1550.8 $m$\AA{} for 2M03455048+2352262, in contrast with 803.2 $m$\AA{} for 2M03385686+2434112. The lower $T_{\rm eff}$ star 2M03455048+2352262 presents a slightly lower
pseudo-continuum of the line $\lambda$15748.988\AA{}  compared to the warmer star 2M03385686+2434112.

\subsubsection*{\textbf{Departures from LTE and use of a different radiative transfer code}}

The observed abundance trend for certain spectral lines may result from several factors, one of which is deviations from the local thermodynamical equilibrium.
As discussed in Section \ref{sec:NLTE}, the majority of lines analyzed in this study showed negligible deviations from LTE ($\sim$0.01 dex for most),
and these cannot explain the spurious abundance trends observed for certain lines.
Manganese lines exhibited more significant departures from LTE (around 0.10 dex), however, such  deviations remained roughly constant across the $T_{\rm eff}$ range of our Pleiades sample. 
Finally, some abundance trends might be attributed to the need for a more comprehensive abundance analysis, combining non-LTE and 3D effects. However, these corrections are expected to be minimal for G and K-dwarfs (\citealt{Osorio2020, Asplund2021}, and references therein).

To investigate if a different radiative transfer code and model atmosphere grid would predict different profiles for the Mg I lines, we derived Mg abundances for ten representative stars covering a range in effective temperature (2M03455048+2352262, 2M03513903+2245010, 2M03450326+2350219, 2M03454184+2425534, 2M03502089+2428003, 2M03483451+2326053, 2M03462735+2508080, 2M03462777+2335337, 2M03491230+2313421, 2M03462862+2445323) using the 1D LTE radiative transfer code MOOG (\citealt{Sneden2012}) with Kurucz atmospheric models (\citealt{Kurucz}) and Turbospectrum with Kurucz atmospheric models. 
The MOOG/Kurucz derived Mg abundances remained similar to the ones from this study, which were derived with Turbospectrum/MARCS, and, in particular, the strong Mg I lines displayed similar trends of A(Mg) versus $T_{\rm eff}$ as before, with the K dwarfs having significantly lower abundances than F and G dwarfs.

\subsubsection*{\textbf{Systematic uncertainties in the adopted parameters}}
Systematic errors in stellar parameters often cause trends in abundance results. We used the sensitivity table for abundance changes due to uncertainties in the $T_{\rm eff}$ and log $g$ to assess the parameter changes needed to align the Mg abundances from the Mg I lines $\lambda$15740.716 \AA{}, $\lambda$15748.988 \AA{}, and $\lambda$15765.842 \AA{} of K dwarfs with those of F dwarfs. Our analysis indicated that achieving similar Mg abundances would require reducing $T_{\rm eff}$ by $\sim$900 K and log $g$ by $\sim$0.6 dex. Such systematic changes are completely out of the uncertainties in the stellar parameters, and we conclude that this is not a viable solution to the abundance trend problem.

\subsubsection*{\textbf{Extra broadening due to Zeeman splitting}}

Young stars, such as those members of the Pleiades open cluster, are known to be more active and have magnetic fields.
The influence of stellar activity or magnetic fields in these stars could introduce extra broadening in those line profiles corresponding to transitions that are sensitive to magnetic fields (see \citealt{Barrado2001,Wanderley2024}).
In this regard, we compiled Landé $g$-factors 
for all lines analyzed in this work to investigate potential correlations between the observed abundance and Landé factors. Our analysis found no correlation between the Landé $g$ values and the observed abundance trends, suggesting that the abundance trends are not related to the effects of the stellar magnetic fields.

\subsubsection*{\textbf{Changing the broadening parameters in the line list}}
Under the reasonable assumption that the chemical abundances in open clusters are homogeneous, we adopt the Mg abundance obtained for a solar-type star from our sample as representative of the Mg abundance for the Pleiades cluster. This choice is made because the Sun was used as the benchmark star to fine-tune damping constants for strong lines in the APOGEE line list, as discussed in \cite{Smith2021}. 

We analyzed the spectrum of a K-dwarf ($T_{\rm eff}$ $\sim$ 4500 K) from our sample, and, adopting the reference Mg abundance from the solar-type star, we adjusted the van der Waals damping values for the three strong Mg I lines at $\lambda$15740.716 \AA, $\lambda$15748.988 \AA, and $\lambda$15765.842 from their default values in the line list by -0.52, -0.47, and -0.68, respectively. These adjustments in the damping led to sharper line profiles, and we note that in order to achieve good fits between observed and synthetic spectra, small additional adjustments in the adopted Gaussian broadening representing the APOGEE spectrograph LSF were required.
%adjusted the van der Waals collisional damping constant until  strong lines in K dwarfs ($T_{\rm eff}$ $\sim$ 4500K) to agree with those of F dwarfs by adjusting the damping values in the line list. 
%The damping used in this work adopted from the APOGEE line list is %(for most of the transitions) 
%the logarithm of the van der Waals collisional damping divided by the hydrogen number density (\citealt{Shetrone2015,Smith2021}). 
%we adjusted the damping values for the Mg I lines at $\lambda$15740.716 \AA, $\lambda$15748.988 \AA, and $\lambda$15765.842 \AA to -0.52, -0.47, and -0.68, from its default value, respectively. This adjustment led to sharper line profiles, and, to achieve a good fit between observed and synthetic spectra, required additional adjustments in the Gaussian broadening. %of 200 $m$\AA{} 
We did the same test for an F-dwarf ($T_{\rm eff} \approx$ 6500 K) from our sample, and this resulted in small positive changes of +0.14, 0.20, and 0.13 in the damping values of the Mg I lines at $\lambda$15740.716 \AA, $\lambda$15748.988 \AA, and $\lambda$15765.842 \AA, respectively. 
As previously discussed, it is clear that the stronger the spectral line, the larger the discrepancy in the modeled line profiles in comparison with the observations if the elemental abundance is kept constant, and consequently, when allowing the abundance to vary, the more pronounced is the trend with $T_{\rm eff}$. 
%likely due to increased electron pressure in these intense lines of K-dwarfs.

%--------------------------------------------
\begin{figure}
  \centering
  \subfloat{\includegraphics[width=0.49\textwidth]{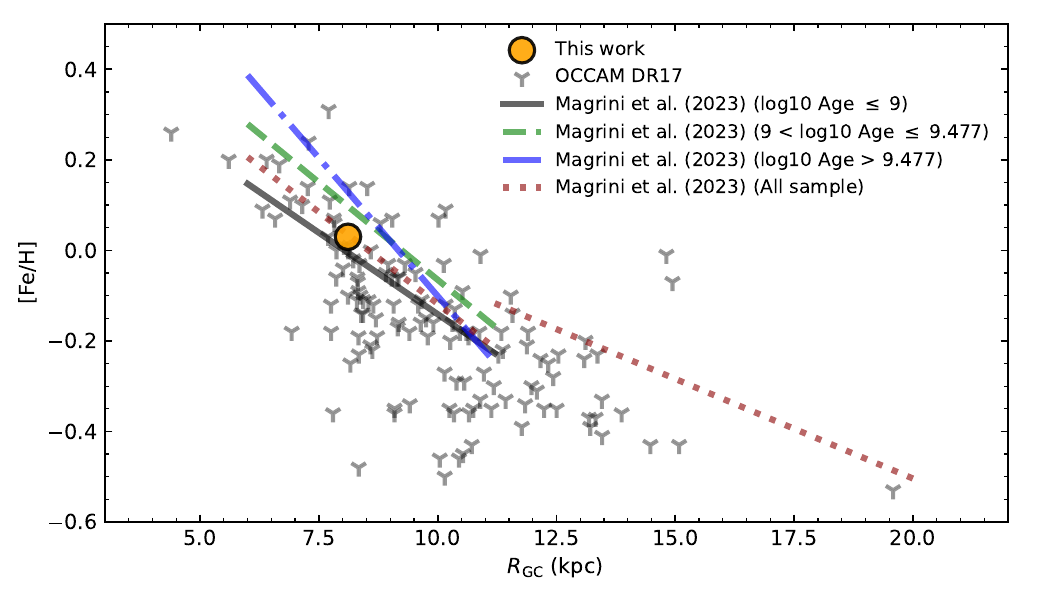}}
  \caption{Stellar metallicity as a function of galactocentric distance (kpc). Our metallicity result for the Pleiades (filled orange circle) plotted along with the OCCAM results, shown as grey symbols, with the different lines representing abundance gradient slopes for different ages from Magrini et al. (2023). 
  }
  \label{fig:gradients}
\end{figure}
%--------------------------------------------

\subsection{\textbf{Limits on atomic diffusion in the Pleiades stars}} \label{sec:difusion}

Atomic diffusion is a significant physical process, theoretically described (\citealt{Michaud2015}), occurring in all stars but more dominant in certain evolutionary stages. Generally, atomic diffusion competes with other hydrodynamic transport processes that can neutralize its effect, with convection being the most significant. Typically, the timescale of convection is orders of magnitude shorter than that of diffusion, thereby significantly diminishing the diffusion effect in convective regions.
The condition for efficient diffusion is the absence of other significant transport processes, such as convection, turbulent motions, radiative winds, accretion, and mass loss. For instance, diffusion is expected to operate efficiently in the radiative zones of solar-type or warmer main-sequence stars, where gravitational settling is effective due to reduced convective motion. However, when solar-type stars evolve into the red giant stage, the effects of diffusion become much less significant (\citealt{Michaud2015}), primarily because the convective zone significantly expands. 

It is expected that the young Pleiades open cluster members do not exhibit large atomic diffusion effects (\citealt{Dotter2017}). The MIST isochrone model, which incorporates atomic diffusion for solar metallicity at the age of 112 Myr (\citealt{Choi2016}), predicts an iron abundance depletion of approximately [Fe/H] = 0.02 dex at $T_{\rm eff}$ around 6500K. Although such a small abundance depletion would not be measurable in this study, given our internal iron abundance uncertainties of $\sim$ 0.05 - 0.06 dex, the few warmer F stars in our sample ($T_{\rm eff}$ $\sim$ 6500K) do not show any statistically significant evidence of depletion beyond the estimated uncertainties (see Figure \ref{fig:teffxelements}). The absence of detectable diffusion beyond $\sim$ 0.05 dex is overall in line with the model predictions.

\subsection{\textbf{The results for the Pleiades in the context of Galactic metallicity gradients}}  \label{sec:gradients}

Open clusters are important probes in defining the metallicity gradients of the Galactic thin disk  (e.g., \citealt{Frinchaboy2013,Cunha2016,Casamiquela2021,Spina2022}).
In Figure \ref{fig:gradients}, we present the galactocentric distance (in kpc) versus metallicity for various galactic open clusters (grey symbols), based on data from the OCCAM survey (\citealt{Myers2022}); the mean metallicity obtained in this study for the Pleiades stars is shown as the filled orange circle. To compare the position of the Pleiades open cluster with metallicity gradients from the literature, we show the results from \cite{Magrini2023}, which come from the open cluster sample in the Gaia-ESO survey. Different gradients in that study were obtained for open clusters in different age bins: the black line represents clusters younger than 1 Gyr, the green-dashed line open clusters having ages between 1 and 3 Gyr, and the dot-dashed blue line represents ages higher than 3 Gyr. The dotted brown line illustrates the gradient for all the open cluster samples from \cite{Magrini2023} ([Fe/H] = -0.081$\times$$R_{\rm GC}$ + 0.692 for $R_{\rm GC}$ $\leq$ 11.2 kpc and [Fe/H] = -0.044$\times$$R_{\rm GC}$ + 0.376 for $R_{\rm GC}$ $>$ 11.2 kpc). 
Notably, our result for the Pleiades is in line with the overall position of the OCCAM open clusters in a similar galactocentric distance. It intersects the overall regression line (brown dashed line), falling slightly above the line representing the gradients for clusters younger than 1 Gyr (black line).

\section{Conclusions}  \label{sec:conclusion}

Open clusters serve as essential astrophysical laboratories for examining physical processes during stellar evolution. They provide unique opportunities to validate methodologies for determining stellar abundances across stars with varied masses, effective temperatures, and surface gravities. Analyzing how stellar abundances correlate with stellar parameters and evolutionary stages can reveal systematic errors in abundance determinations or reveal signatures of abundance variations.

This paper presents a comprehensive study of the abundances of C, Na, Mg, Al, Si, K, Ca, Ti, V, Cr, Mn, and Fe in 80 F, G, and K main-sequence dwarf members of the Pleiades open cluster, utilizing high-resolution APOGEE spectra. Our sample spans an effective temperature range of $\sim$4500 to 6850 K. Despite being one of the most extensively studied open clusters, the Pleiades previously lacked a detailed abundance analysis of many elements covering a wide range of stellar masses as presented here.

The mean metallicity obtained for the studied Pleiades sample is roughly solar, [Fe/H] = 0.03 $\pm$ 0.04 dex. The mean element-over-iron abundance ratios derived for the different nucleosynthetic families are as follows: [$\alpha$/Fe] = 0.01 $\pm$ 0.05, [odd-z/Fe] = -0.04 $\pm$ 0.08, and [iron peak/Fe] = -0.02 $\pm$ 0.08. All these abundance ratios %are consistent with results from the literature for the cluster (\citealt{Soderblom2009, Takeda2016, Spina2018}) and 
agree with expectations for the chemistry of a roughly solar metallicity open cluster from the solar neighborhood. In addition, the mean metallicity obtained for the Pleiades are in line with metallicity gradients from open clusters in the literature.

%Our derived abundances show good agreement with those reported in the literature. Specifically, when compared with \cite{Spina2018}, we observed a close match in metallicity, with [Fe/H] $\delta$ (This work - \citealt{Spina2018}) = 0.05 $\pm$ 0.05 dex. The abundances displaying higher discrepancies were Na, Al, and Cr; however, considering the uncertainties in both studies, the results remain statistically similar.
%\textbf{The comparison with APGOEE DR17 also shows good agreement. Exceptions are for Na and V, where the standard deviation of the mean for both [Na/H] and [V/H] in DR17 is 0.23 and 0.10, respectively, while for this work, we obtain 0.04 and 0.09.}

The detailed line-by-line modeling presented for the studied Pleiades dwarfs, which cover a wide range in effective temperature of more than 2000 K, was an effective tool in revealing inadequacies in the modeling of strong lines in the APOGEE region when using the current abundance analysis codes and the APOGEE line list.  Allowing the elemental abundances to vary in order to compensate for mismatches between synthesis and observations resulted in derived abundances having strong trends with the effective temperature, for example, for the three strong Mg I lines at $\lambda$15740.716\AA, $\lambda$15748.988\AA, and $\lambda$15765.842\AA. 

One important result of this paper is the compilation of selected spectral lines (Table \ref{tab:apendix}), which are deemed to be good abundance indicators and to provide abundances that are nearly independent of the effective temperature for F, G, and K dwarfs.
The APOGEE abundance pipeline ASPCAP derives abundances by fitting, at the same time, all selected lines available for a specific element, and as thus, is also affected by the inadequate modeling of strong lines in the APOGEE spectra, in particular for K dwarfs.  

%\textbf{A detailed examination of abundances derived for each individual spectral line revealed trends as a function of effective temperature for certain spectral lines, which were identified as} the stronger lines, such as certain Mg \textbf{I} or Al \textbf{I} lines.  Such lines were excluded (\textbf{Table \ref{tab:apendix}}) from the final abundance results reported in Table \ref{tab:abundances}. 
%We applied non-LTE corrections based on literature results and found that these deviations are minor (less than 0.01 dex for most elements), which does not fully explain the observed trend.
We performed several tests on the strong Mg I lines to probe the observed abundance trends with the effective temperature, such as using different radiative transfer codes or stellar atmosphere models, evaluating if realistic changes in the stellar parameters would erase the abundance trends, investigating non-LTE effects for all studied lines, examining possible trends of abundances with Land\'e g factors, and adjusting the van der Waals damping constants while keeping the Mg abundance constant. Only the latter could reproduce the observed line profiles, suggesting inadequacies in the spectral line broadening, especially for the cooler K dwarfs, as a possible cause of the spurious trends. However, further work involving K dwarf members from other open clusters is needed in order to explore this possibility carefully.

\section*{Acknowledgements}
We thank the referee for suggestions that improved the paper significantly.
D.S. and R.V. thank the National Council for Scientific and Technological Development – CNPq process No. 404056/2021-0.
K.C. acknowledges support from the National Science Foundation through NSF grant No. AST-2206543.
F.W. acknowledges support from fellowship by Coordena\c c\~ao de Ensino Superior - CAPES. 
RG gratefully acknowledges the grants support provided by ANID Fondecyt Postdoc No. 3230001 and FAPERJ under the PDR-10 grant number E26-205.964/2022.
SD acknowledges CNPq/MCTI for grant 306859/2022-0.

Funding for the Sloan Digital Sky Survey IV has been provided by the Alfred P. Sloan Foundation, the U.S. Department of Energy Office of Science, and the Participating Institutions. SDSS-IV acknowledges support and resources from the Center for High-Performance Computing at the University of Utah. The SDSS website is www.sdss.org.
SDSS-IV is managed by the Astrophysical Research consortium for the Participating Institutions of the SDSS Collaboration including the Brazilian Participation Group, the Carnegie Institution for Science, Carnegie Mellon University, the Chilean Participation Group, the French Participation Group, Harvard-Smithsonian Center for Astrophysics, Instituto de Astrof\'isica de Canarias, The Johns Hopkins University, 
Kavli Institute for the Physics and Mathematics of the Universe (IPMU) /  University of Tokyo, Lawrence Berkeley National Laboratory, Leibniz Institut f\"ur Astrophysik Potsdam (AIP),  Max-Planck-Institut f\"ur Astronomie (MPIA Heidelberg), Max-Planck-Institut f\"ur Astrophysik (MPA Garching), Max-Planck-Institut f\"ur Extraterrestrische Physik (MPE), National Astronomical Observatory of China, New Mexico State University, New York University, University of Notre Dame, Observat\'orio Nacional / MCTI, The Ohio State University, Pennsylvania State University, Shanghai Astronomical Observatory, United Kingdom Participation Group,
Universidad Nacional Aut\'onoma de M\'exico, University of Arizona, University of Colorado Boulder, University of Oxford, University of Portsmouth, University of Utah, University of Virginia, University of Washington, University of Wisconsin, Vanderbilt University, and Yale University.

{\it Facilities: {Sloan}}.

Software: BACCHUS (\citealt{Masseron2016}), Turbospectrum (\citealt{Alvarez1998}; \citealt{Plez2012}; \href{https://github.com/bertrandplez/Turbospectrum2019}{https://github.com/bertrandplez/Turbospectrum2019}).

%%%%%%%%%%%%%%%%%%%%%%%%%%%%%%%%%%%%%%%%%%%%%%%%%%
\section*{Data Availability}

We utilize data from the final public data release of the SDSS-IV/APOGEE, specifically DR17. The data are publicly available in \href{https://www.sdss4.org/dr17/irspec/spectro_data/}{https://www.sdss4.org/dr17/irspec/spectro\_data/}.

% The inclusion of a Data Availability Statement is a requirement for articles published in MNRAS. Data Availability Statements provide a standardised format for readers to understand the availability of data underlying the research results described in the article. The statement may refer to original data generated in the course of the study or to third-party data analysed in the article. The statement should describe and provide means of access, where possible, by linking to the data or providing the required accession numbers for the relevant databases or DOIs.



%\bibliographystyle{mnras}


%%%%%%%%%%%%%%%%%%%%%%%%%%%%%%%%%%%%%%%%%%%%%%%%%%

%%%%%%%%%%%%%%%%% APPENDICES %%%%%%%%%%%%%%%%%%%%%

\onecolumn
\appendix
\section{Spectral lines adopted or removed from this work}

% \onecolumn
\renewcommand{\arraystretch}{0.9}
\begin{longtable}{
>{\columncolor[HTML]{FFFFFF}}c 
>{\columncolor[HTML]{FFFFFF}}c 
>{\columncolor[HTML]{FFFFFF}}c }
\caption{Adopted and Unadopted Lines} \\
\hline
Element & Adopted Lines (\AA) & Excluded Lines (\AA) \\ \hline
Fe I & 15395.1 & 15207.5 \\
    & 15588.1 & 15245.0 \\
    & 15592.2 & 15294.6 \\
    & 15604.2 & 15335.2 \\
    & 15621.7 & 15501.1 \\
    & 15632.1 & 15662.0 \\
    & 15686.3 & 15677.5 \\
    & 15691.9 & 15723.6 \\
    & 15904.4 & 15774.1 \\
    & 15920.7 & 15868.6 \\
    & 15967.7 & 15892.6 \\
    & 16006.8 & 15911.3 \\
    & 16071.5 & 15964.9 \\
    & 16175.0 & 15980.7 \\
    & 16179.5 & 16037.8 \\
    & 16180.9 & 16040.7 \\
    & 16185.9 & 16042.7 \\
    & 16195.1 & 16075.9 \\
    & 16204.2 & 16125.9 \\
    & 16207.7 & 16165.0 \\
    & 16213.5 & 16231.7 \\
    & 16225.6 & 16517.2 \\
    & 16284.8 & 16524.5 \\
    & 16316.4 & 16561.8 \\
    & 16394.4 & 16645.9 \\
    & 16398.2 & \\
    & 16486.6 & \\
    & 16522.1 & \\
    & 16541.6 & \\
    & 16665.5 & \\
    & 16753.1 & \\ \hline
C I & 15784.7 & 16004.9 \\
    &  & 16021.7 \\ \hline
Na I & 16373.9 & \\
    & 16388.9 & \\\hline
Mg I & 15879.5 & 15740.7 \\
    & 15886.2 & 15749.0 \\
    & 15954.5 & 15765.8 \\ \hline
Al I & 16763.4 & 16719.0 \\
    &  & 16750.6 \\ \hline
Si I & 15361.2 & 15888.4 \\
    & 15376.8 & 15960.1 \\
    & 16215.7 & 16060.0 \\
    & 16680.8 & 16094.8 \\
    & 16828.2 & \\ \hline
K I & 15163.1 & \\
    & 15168.4 & \\ \hline
Ca I & 16136.8 & \\
    & 16150.8 & \\
    & 16155.2 & \\
    & 16157.4 & \\ \hline
Ti I & 15602.8 & 15334.8 \\
    & 15699.0 & 15543.8 \\
    & 16635.2 & 15715.6 \\ \hline
V I & 15924.0 & \\ \hline
Cr I & 15680.1 & \\ \hline
Mn I & 15159.0 & 15217.0 \\
    & 15262.0 & \\
\hline
\label{tab:apendix}
\end{longtable}

% If you want to present additional material which would interrupt the flow of the main paper, it can be placed in an Appendix which appears after the list of references.

%%%%%%%%%%%%%%%%%%%%%%%%%%%%%%%%%%%%%%%%%%%%%%%%%%

% Don't change these lines
\bsp	% typesetting comment
\label{lastpage}
\end{document}